\documentclass[10pt,letterpaper]{article}
\usepackage[top=0.85in,left=2.75in,footskip=0.75in]{geometry}\usepackage{graphicx}  
\usepackage{changepage}

\usepackage[utf8x]{inputenc}

\usepackage{textcomp,marvosym}

\usepackage{fixltx2e}

\usepackage{amsmath,amssymb}

\usepackage{cite}

\usepackage{nameref,hyperref}

\usepackage[right]{lineno}

\usepackage{microtype}
\DisableLigatures[f]{encoding = *, family = * }

\raggedright
\usepackage[aboveskip=1pt,labelfont=bf,labelsep=period,justification=raggedright,singlelinecheck=off]{caption}

\makeatletter
\renewcommand{\@biblabel}[1]{\quad#1.}
\makeatother

\date{}

\usepackage{lastpage,fancyhdr,graphicx}
\usepackage{epstopdf}
\fancyheadoffset[L]{2.25in}
\fancyfootoffset[L]{2.25in}

\begin{document}
\vspace*{0.2in}

\begin{flushleft}
{\Large
\textbf\newline{Dynamics of Adaptive Immunity Against Phage in Bacterial Populations} 
}
\newline
\\
Serena Bradde\textsuperscript{1,2\Yinyang},
Marija Vucelja\textsuperscript{3,4\Yinyang},
Tiberiu Te\c sileanu\textsuperscript{1,2},
Vijay Balasubramanian\textsuperscript{1,2*}
\\
\bigskip
\textbf{1} Initiative for the Theoretical Sciences, The Graduate Center, CUNY, New York, NY 10016 \\
\textbf{2} David Rittenhouse Laboratories, University of Pennsylvania, Philadelphia, PA 19104 \\
\textbf{3} Center for Studies in Physics and Biology, The Rockefeller University, New York, NY 10065 \\
\textbf{4} Department of Physics, University of Virginia, Charlottesville, VA 22904 \\
\bigskip

\Yinyang These authors contributed equally to this work.

*vijay@physics.upenn.edu
\end{flushleft}

\section*{Abstract}
The CRISPR (clustered regularly interspaced short palindromic repeats) mechanism allows bacteria to adaptively defend against phages by acquiring short genomic sequences (spacers) that target specific sequences in the viral genome.   We propose a population dynamical model where immunity can be both acquired and lost.   The model predicts regimes where bacterial and phage populations can co-exist,  others where the populations exhibit damped oscillations, and still others where one population is driven to extinction. Our model considers two key parameters: (1) ease of acquisition and (2) spacer effectiveness in conferring immunity.    Analytical calculations and numerical simulations show that  if spacers differ mainly in ease of acquisition, or if the probability of acquiring them is sufficiently high, bacteria develop a  diverse population of spacers.  On the other hand, if spacers differ mainly in their effectiveness, their final distribution will be highly peaked, akin to a ``winner-take-all'' scenario, leading to a specialized spacer distribution.    Bacteria can interpolate between these  limiting behaviors by actively tuning their overall acquisition probability.


\section*{Author Summary}

The CRISPR system in bacteria and archaea provides adaptive immunity by incorporating foreign DNA (spacers) into the genome, and later targeting DNA sequences that match these spacers.  The way in which bacteria choose  spacer sequences from a clonal phage population is not understood.  Our model considers  competing effects of ease of acquisition and effectiveness against infections in shaping the spacer distribution.  The model suggests that a diverse spacer population results when the acquisition rate is high, or when spacers are similarly effective.  At moderate acquisition rates, the spacer distribution becomes highly sensitive to spacer effectiveness.  There is a rich landscape of behaviors including bacteria-phage coexistence and oscillations in the populations.

\section*{Introduction}
\begin{figure}[!h]
	\center\includegraphics[width=3.5in]{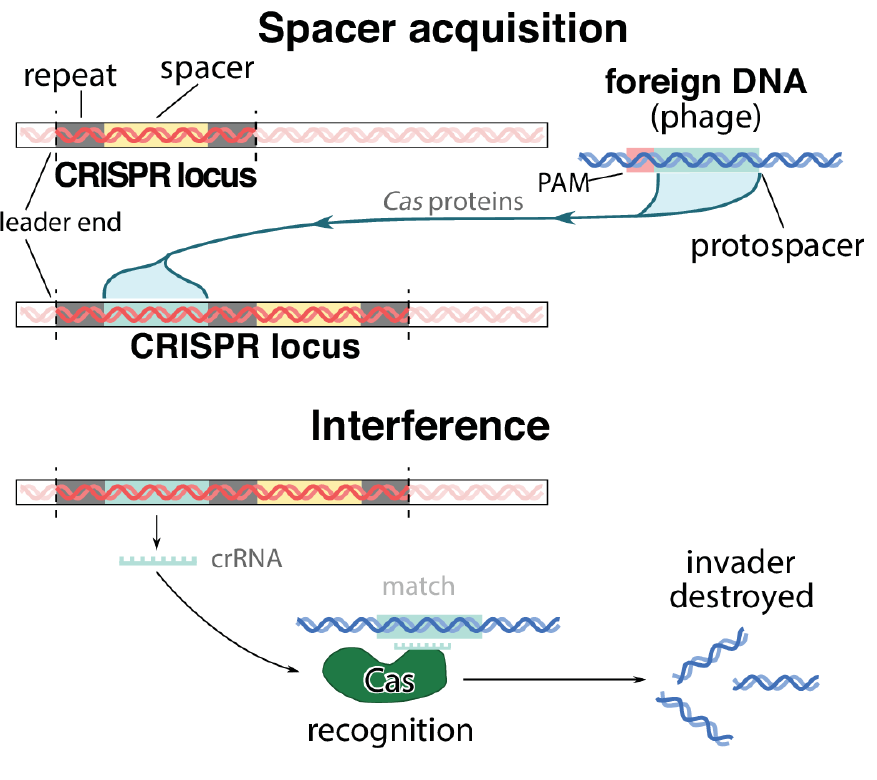}
	\smallskip
	\caption{Schematic of the CRISPR acquisition and interference mechanism. PAM stands for protospacer adjacent motif, a short sequence necessary for protospacer recognition by the \textit{Cas} proteins.\label{fig:crispr_intro}}
\end{figure}

Bacteria and archaea can combat viral infections using innate mechanisms (\textit{e.g.}, abortive infection, surface exclusion and restriction modification systems) that are not specific to particular threats~\cite{Labrie2010, Thomas2005, Bikard2012a}. Some species also exhibit an adaptive immune system based on CRISPR (clustered regularly interspaced short palindromic repeats) interference, which allows bacteria to specifically target and cleave exogenous genetic material from previously encountered phages and other genetic elements~\cite{Barrangou2007,  Marraffini2008, Heler2014, Brodt2011}.  The system works by incorporating short (30--70 bp) sequences, dubbed ``spacers'', into the bacterial genome, in between repeated CRISPR elements (Fig.~\ref{fig:crispr_intro}). The spacers are acquired from the ``protospacer'' regions in the genome of infecting phage.    CRISPR Type I and II  require the presence of a ``protospacer adjacent motif'' (PAM) upstream of a protospacer for recognition by the CRISPR proteins \cite{Makarova2011}.   The PAM sequence is thought to play a role in the avoidance of auto-immune targeting \cite{Marraffini2010}.   While the PAM and the first few nucleotides of the protospacer (the ``seed'' region) need to match almost perfectly for CRISPR interference \cite{Heler2014}, there is significant tolerance to mutations in the rest of the spacer~\cite{Semenova2011}.

Over the whole viral genome, there can be  tens or hundreds of protospacers, and the way in which the CRISPR acquisition mechanism selects between these is not fully understood \cite{Fineran2012}. Experiments show that after several hours of exposure of bacteria to phage, different spacers occur with different frequencies, with a handful being much more abundant~\cite{Paez-Espino2013}. Importantly, many of the highly abundant spacers recur during repetition of the experiments, suggesting that their over-representation is not simply the result of amplification of spacers that are randomly acquired at the early stage of infection.   There are three main possible sources of selective pressure on spacers. One is a bias in acquisition that may arise either when some protospacers are easier to acquire by the CRISPR proteins than others~\cite{Levy2015}, or when some protospacers are more conserved in the viral population, and thus more abundant and more likely to be acquired. Another possible source of selective pressure is that some spacers might be more effective than others at clearing viral infections and so provide a selective advantage for the host~\cite{Barrangou2007, Semenova2011}.  Finally, the acquisition of some spacers might be ``primed'' by the presence of other spacers in the CRISPR locus~\cite{Heler2014,Fineran2012,Datsenko2012,Fineran2014}.

We construct a population dynamical model for bacteria that use CRISPR-based immunity to defend against phage.  Our model predicts that even when dilution is negligible, wild-type and spacer-enhanced bacteria can co-exist with phage, provided there  is spacer loss.  Previous Lotka-Volterra-like ecological models have demonstrated a mechanism for coexistence between three species with bounded populations, but, unlike the scenario we describe, they required dilution and significant  differences in the growth rates of the two prey species \cite{Levin1977}.   
  To understand the factors that affect spacer diversity, we compare two scenarios:  (a)  different spacers are acquired at different rates; (b)  different spacers provide different advantages, \textit{e.g.,} in growth rate or survival rate, to the host.      
 We derive analytical results for the spacer distribution that is reached at late times, and show that the spacer-effectiveness model favors a peaked distribution of spacers while the spacer-acquisition model favors a more diverse distribution.  Higher rates of spacer acquisition also lead to higher diversity.    We expect that greater spacer diversity will be important for defending against  a mutating phage landscape, while a peaked spacer distribution will confer stronger immunity against a specific threat.      Our model predicts that bacteria can negotiate this tradeoff by controlling the overall rate at which spacers are acquired, \textit{i.e.,} by modifying the expression of the \textit{Cas} proteins, necessary for acquisition \cite{Heler2014}.

\section*{Model}

We consider bacteria that start with a CRISPR cassette containing no spacers, a scenario that has been proven functional \textit{in vivo}~\cite{Yosef2012}. We focus on the early dynamics of the bacterial population after being infected with phage in which each bacterial cell acquires at most one spacer. Experiments suggest that this  scenario may be appropriate for bacteria-phage interactions lasting about a day, which allows most of the bacterial population to become immune to the infecting phage, but is not enough time for viral escapers that can avoid the bacterial defenses to become abundant~\cite{Levin2013, Paez-Espino2013}. In the absence of escapers, the acquisition of new spacers against the same virus is slow~\cite{Datsenko2012}, extending the duration for which our single spacer approximation is valid.
As time goes by, the virus will mutate and the bacteria need to acquire new spacers to keep up with the mutants; we leave the study of this co-evolution to future work, and focus here on the early dynamics of spacer acquisition. 

Even if each bacterial cell only has time to acquire at most one spacer, the population as a whole will contain a diverse spacer repertoire~\cite{Paez-Espino2013, Childs2014, Andersson2008}. Here we propose a model of bacteria-phage dynamics to understand the distribution of spacers in the population.  As a warmup, we  first study the case where the virus contains only a single protospacer, then we generalize the model to the case of many protospacers where acquisition probability and effectiveness can depend on the type.

\subsection*{One spacer type}
\begin{figure}[!h]
	\center\includegraphics[width=5in]{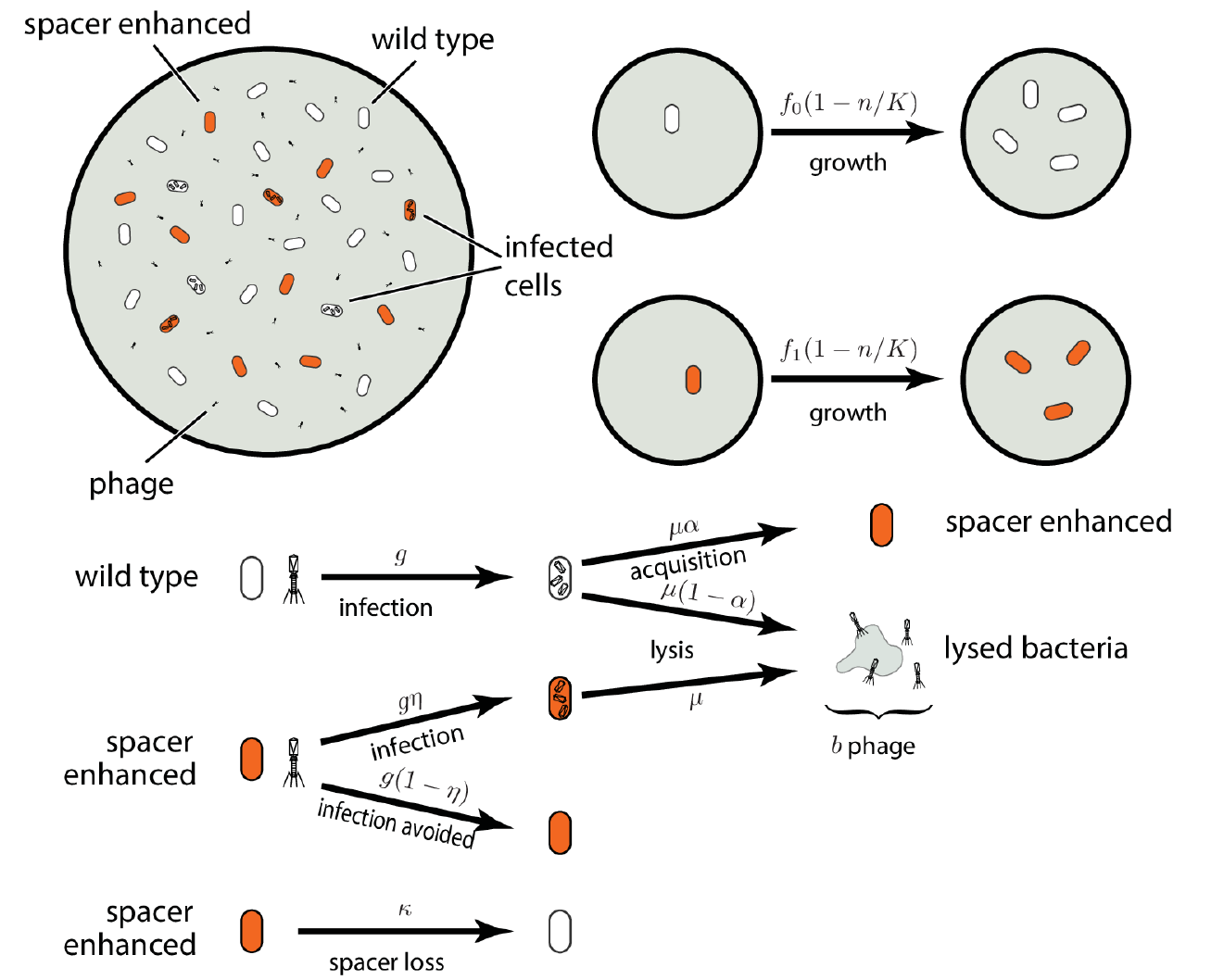}
	\bigskip
	\caption{Model of bacteria and phage dynamics.  Bacteria are either wild type or spacer enhanced, grow  at different rates $f_0$ and $f_1$ and can be infected by phage with rates $g$ and $\eta g$. Spacers can be acquired during infection with a probability $\alpha$ and spacers are lost at a  rate $\kappa$. \label{fig:model_in_fig}}
\end{figure}

To set the stage, we will first introduce the dynamics of a model where viruses  present a single protospacer.   In this case, all immune bacteria have the same spacer.  We will assume logistic growth of the bacteria~\cite{Renshaw1991}.   The relevant processes are sketched in Fig.~\ref{fig:model_in_fig} 
and, assuming a well-mixed population, can be translated into a set of ordinary differential equations:
\begin{eqnarray}\label{eq:single_proto_spacer}
	\dot n_0 &=& f_0 \left(1- \frac{n}{K}\right) n_0 + \kappa n_1 - g v n_0 \,, \nonumber\\
	\dot n_1 &=& f_1 \left(1-\frac{n}{K}\right) n_1 - \kappa n_1   - \eta g v n_1 + \alpha \mu I_0 \,,\nonumber \\
	\dot I_0 &= &g v n_0 - \mu I_0\,,\nonumber \\
	\dot I_1 &= &\eta g v n_1 - \mu I_1\,,\nonumber\\
	\dot v \ &=& b \left(1-\alpha\right) \mu I_0 + b \mu I_1 -g v (n_0+n_1)\, .
\end{eqnarray}
Here the dot represents the derivative with respect to time, $n_0$ is the number of ``wild type'' bacteria that do not contain any spacers, $n_1$ is the number of ``spacer enhanced'' bacteria that have acquired the spacer,  $I_0$ is the number of wild-type infected bacteria, and $I_1$ is the number of spacer enhanced but infected bacteria (which is possible because spacers do not provide perfect immunity).    The sizes of the bacterial and phage populations are 
$$n=n_0+n_1+I_0+I_1$$
and $v$ respectively.

The first term in the first two equations in~\eqref{eq:single_proto_spacer} describes logistic growth of the bacteria with maximum growth rates $f_i$ and a carrying capacity $K$.  These equations allow for the possibility that spacer enhanced bacteria may grow at a different rate than the wild type because  of possible spacer toxicity due to auto-immune interactions or due to increased metabolic rate arising from expression of CRISPR (\textit{Cas}) proteins and/or CRISPR RNA \mbox{(crRNA)}.   However,  there is evidence~\cite{Paez-Espino2013,Jiang2013}  that these growth rate differences are small so that  $r=f_1/ f_0\approx 1$.  
We also assume that spacers can be lost at a rate $\kappa$  (second term in the first and second equations) allowing bacteria to revert to wild type~\cite{Jiang2013, Tyson2008,Iranzo2013}.
Bacteria become infected with different rates depending on their type---wild type are always infected if they encounter phage, but spacer enhanced bacteria may evade infection.  Taking $g$ to be the encounter rate, wild type are infected at a rate $g$ while spacer enhanced bacteria are infected at a rate $\eta g$ where $\eta < 1$ (third terms of the first and second equations).  We can think of $\eta$ as a ``failure probability'' of the spacer as a defense mechanism, or alternatively, of $1-\eta$ as a measure of the ``effectiveness'' of the spacer against infections. Finally, some infected wild-type bacteria survive and acquire a spacer with probability $\alpha$ (last term in the second equation).    We can imagine that this acquisition occurs in the course of an infection that is unsuccessful because the phage is ineffective or because of innate immune mechanisms, while nevertheless allowing the bacterial cell access to genetic material of the phage.   We are neglecting the possibility that spacers might also be acquired via horizontal gene transfer without an infection.
 
The dynamics of the infected bacteria is given in the third and fourth equations in \eqref{eq:single_proto_spacer}.   We assume that infected bacteria do not divide. So the number of infected bacteria grows only because of new infections (first terms in the equations),  and declines due to lysis or successful defense followed by acquisition of spacers (second term). The lysis rate $\mu$ depends on  properties of the phage including the burst factor $b$ (\textit{i.e.,} the number of viral particles produced before lysis).    More specifically, there is a  delay between infection and lysis because it takes some time for the virus to reproduce.   We are approximating this delay with a stochastic process following an exponential distribution with timescale $1/\mu$ \cite{Bonachela2014,Weld2004}.

Finally, the last equation describes the dynamics of free phage. The first two terms model  viral replication.   Phage that duplicate in infected bacteria  produce $b$ new copies after cell lysis.   The first term describes this process in infected wild type bacteria that do not acquire a spacer and become immune.  The second term describes the lysis of bacteria that were infected despite having a spacer.  We could imagine that a small number of spacer enhanced bacteria that become infected then become resistant again, perhaps by acquiring a second spacer.   We neglect this because the effect is small for two reasons---acquisition is rare, $\alpha \ll 1$, and because we assume that the spacer is  effective, $\eta \ll 1$, such that $I_1$ is small. The approximation $\eta\ll 1$ is supported by experimental evidence that shows that a single spacer seems often sufficient to provide almost perfect immunity \cite{Barrangou2007}.

For simplicity, our model does not include the effects of natural decay of phage and bacteria as these happen on timescales that are relatively long compared to the dynamics that we are studying.  Likewise, we did not consider the effects of dilution which can happen either in controlled experimental settings like chemostats, or in some kinds of open environments.  In S1 file we show that dilution and natural decay of typical magnitudes do not affect the qualitative character of our results.

We can also write an equation for the total number of bacteria $n$:
\begin{equation}
	\label{eq:single_proto_total}
	\dot n= f_0( n_0 + r n_1) \left(1-\frac{n}{K}\right) -\mu (1-\alpha) I_0 - \mu I_1\,,
\end{equation}
where we used the notation $r=f_1/f_0$.  The total number of bacteria is a useful quantity, since optical density measurements can assess it in real time.

\subsection*{Multiple spacer types}
Typically the genome of a given bacteriophage contains several protospacers as indicated by the occurrence of multiple PAMs.  Even though in the short term each bacterial cell can acquire only one spacer type, at the level of the whole population many types of spacers will be acquired, corresponding to the different viral protospacers. Experiments show that the frequencies with which different spacers occur in the population are highly non-uniform, with a few spacer types dominating~\cite{Paez-Espino2013}. This could happen either because some spacers are easier to acquire than others, or because they are more effective at defending against the phage.

We can generalize the population dynamics in \eqref{eq:single_proto_spacer} to the more general case of $N$ spacer types.
Following experimental evidence~\cite{Jiang2013} we assume that all bacteria, with  or without spacers, grow at similar rates ($f$)---the effect of having different growth rates is analyzed in S1 file.
We take  spacer $i$ to have  acquisition probability $\alpha_i$ and failure probability $\eta_i$. As before, we can alternatively think of $1-\eta_i$ as the effectiveness of the spacer against infection.
The dynamical  equations describing the bacterial and viral populations  become
\begin{align}
\label{multi_acquisition}
\dot n_0 &=  f\left(1- \frac{n}{K}\right) n_0 + \kappa \sum_{i=1}^N n_i - g v n_0\,,\nonumber\\
\dot n_i &=f \left(1-\frac{n}{K}\right) n_i - \kappa n_i  - \eta_i g v n_i + \alpha_i \mu I_0\,,\nonumber \\
\dot I_0 &= g v n_0 - \mu I_0\,,\nonumber\\
\dot I_i &= \eta_i g v n_i - \mu I_i\,,\nonumber\\
\dot v &= b \Bigl(1-\sum_{i=1}^N \alpha_i\Bigr)\mu I_0 + b  \mu \sum_{i=1}^N I_i  - g v \Bigl(n_0 + \sum_{i=1}^N  n_i\Bigr)\,
\end{align}
where  $n_i$ and $I_i$ are the numbers of healthy and infected bacteria with spacer type $i$, and
$\alpha=\sum_{i=1}^N \alpha_i$ is the overall probability of wild type bacteria surviving and acquiring a spacer, since the $\alpha_i$ are the probabilities of disjoint events. This implies that $\alpha < 1$.
The total number of bacteria is governed by the equation
\begin{equation}
	\label{eq:multi_dntotal}
	\dot n = f\Bigl(1-\frac nK\Bigr) \Bigl(n - \sum_{i=0}^{N} I_i\Bigr) - \mu (1-\alpha) I_0 - \mu \sum_{i=1}^N I_i\,.
\end{equation}

\section*{Results}

The two models presented in the previous section can be studied numerically and  analytically. We use the single spacer type model to find conditions under which host--virus coexistence is possible.   Such coexistence has been observed in experiments~\cite{Levin2013} but has only been explained through the introduction of as yet unobserved infection associated enzymes that affect spacer enhanced bacteria~\cite{Levin2013}.   Host-virus coexistence has been shown to occur in classic models with serial dilution \cite{Levin1977}, where a fraction of the bacterial and viral population is periodically removed from the system. Here we show additionally that coexistence is possible without dilution provided bacteria can lose immunity against the virus. We then generalize our results to the  case of many protospacers where we characterize the relative effects of the ease of acquisition and effectiveness on spacer diversity in the bacterial population.

\subsection*{Extinction versus coexistence with one type of spacer}

\begin{figure}
\begin{center}
\includegraphics[width=1.\columnwidth]{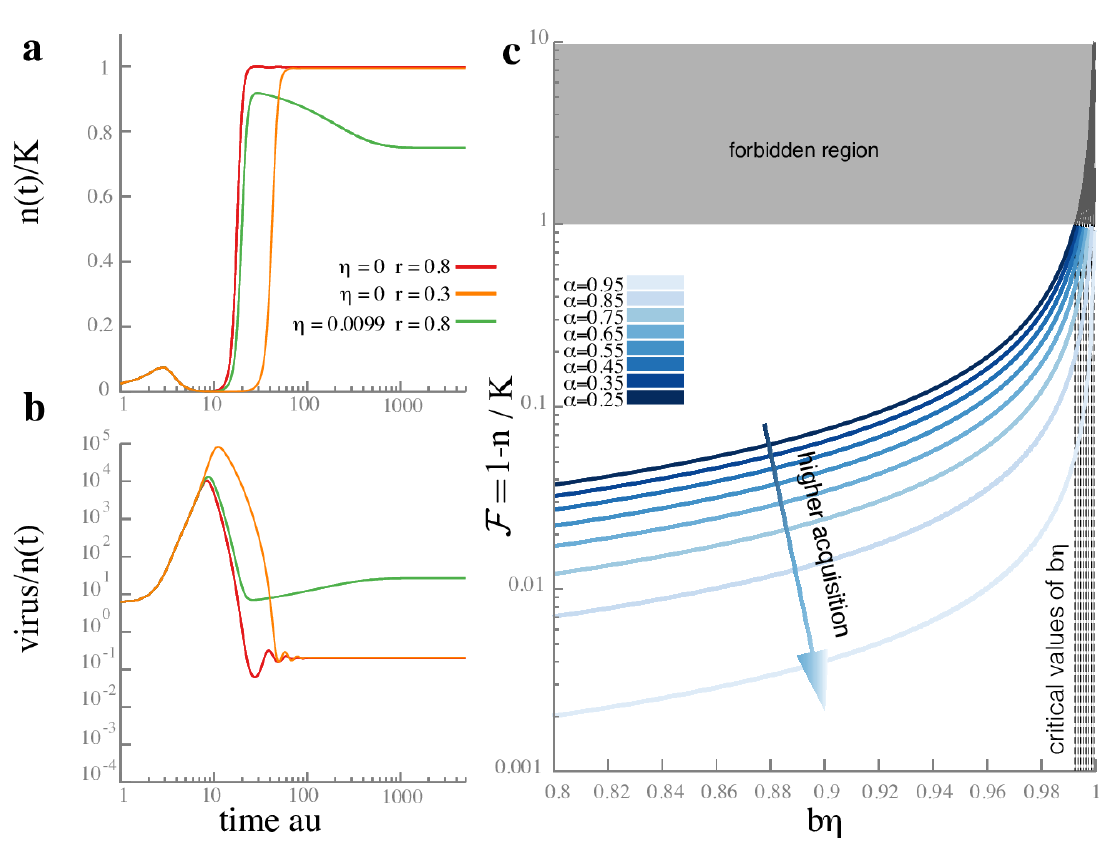}
\end{center}
\caption{Model of bacteria with a single spacer in the presence of lytic phage. (Panel \textbf{a}) shows  the dynamics of the bacterial concentration in units of the carrying capacity $K=10^5$ and (Panel \textbf{b}) shows the dynamics of the phage population. In both panels, time is shown in units of the inverse growth rate of wild type bacteria ($1/f_0$) on a logarithmic scale. Parameters are chosen to illustrate the coexistence phase and damped oscillations in the viral population:  the acquisition probability is $\alpha=10^{-4}$, the burst size upon lysis is $b=100$.  All rates are measured in units of the wild type growth rate $f_0$: the adsorption rate is $g/f_0=10^{-5}$, the lysis rate of infected bacteria is $\mu/f_0=1$, and the spacer loss rate is  $\kappa/f_0= 2 \times 10^{-3} $.  The spacer failure probability ($\eta$) and growth rate ratio $r = f_1/f_0$ are as shown in the legend. The initial bacterial population was all wild type, with a size $n(0) = 1000$, while the initial viral population was $v(0) = 10\,000$.
  The bacterial population has a bottleneck after lysis of the bacteria infected by the initial injection of phage, and then recovers due to CRISPR immunity.  Accordingly, the viral population reaches a peak when the first bacteria burst, and drops after immunity is acquired.  A higher failure probability $\eta$ allows a higher steady state phage population, but oscillations can arise because bacteria can lose spacers (see also S1 file).
 (Panel \textbf{c}) shows the fraction of unused capacity at steady state (Eq.~\eqref{eq:single_proto_f_factor})
 as a function of the product of failure probability and burst size ($\eta b$) for a variety of acquisition probabilities ($\alpha$).
In the plots, the burst  size upon lysis is  $b=100$, the growth rate ratio is $f_1/f_0 = 1$, and the spacer loss rate is $\kappa/f_0=10^{-2} $.
 We see that the fraction of unused capacity diverges as the failure probability approaches the critical value $\eta_c \approx 1/b$ (Eq.~\eqref{eq:constr2}) where CRISPR immunity becomes ineffective.
The fraction of unused capacity decreases linearly with the acquisition probability following Eq.~\eqref{eq:single_proto_f_factor}.
 \label{fig:resulttwobacteria}}
\end{figure}

The numerical solution of the single-spacer population dynamics model is shown in Fig.~\ref{fig:resulttwobacteria}a and Fig.~\ref{fig:resulttwobacteria}b for different parameter choices; more details can be found in S1 file.  In all cases, the bacterial population grows initially because infected bacteria  do not die instantly.  If the viral load is high, most bacteria are quickly infected and growth starts slowing down since infected bacteria cannot duplicate. After a lag of order $1/\mu$, where $\mu$ is the rate at which infected bacteria die, the population declines due to lysis.   If the viral load is low
, the division of healthy bacteria dominates the death of infected ones, until the viral population released by lysis becomes large enough to infect a substantial fraction of the bacteria.

Some infected bacteria acquire the spacer that confers partial immunity from the phage.   During every encounter between a bacterial cell and a virus, there is a probability $\eta$ that the spacer will be ineffective.  Thus the expected increase in the number of viral particles following an encounter is $b \eta - 1$ where $b$ is the viral burst size following lysis of an infected cell. If $\eta>1/b$, the viral growth cannot be stopped by CRISPR immunity and the bacteria are eventually overwhelmed by the infection. Thus whenever the virus has a high burst factor, only a population with an almost perfect spacer (the failure probability $\eta < 1/b \ll 1$) is able to survive infection.

The viral concentration has a more complex dynamics---it typically reaches a maximum, then falls due to CRISPR interference, and starts oscillating at a lower value (Fig.~\ref{fig:resulttwobacteria}b).  The initial rise of the viral population occurs because of successful infections of the wild-type bacteria.   But then, the bacteria which have acquired effective spacers grow exponentially fast, virtually unaffected by the presence of the virus. Since the virus is adsorbed by immune bacteria, but are cleaved by CRISPR and cannot duplicate, the viral population declines exponentially.  However, as the population of spacer-enhanced bacteria rises, so does the population of wild type, because of the constant rate of spacer loss. This  starts a new growth period for the virus, leading to the oscillations seen in simulations.  When spacer effectiveness is low, the virus can still have some success infecting spacer-enhanced bacteria, and the oscillations are damped. It would be interesting to test whether large oscillations in the viral concentration can be seen in experiments to see if these are compatible with measured estimates of the rate of spacer loss $\kappa$ in the context of our model~\cite{Jiang2013, Deveau2008}.

  Varying the growth rate of the bacteria with CRISPR relative to the wild type has a strong effect on the length of the initial lysis phase and the delay before exponential decay of the viral population sets in. In contrast, a lower effectiveness of the CRISPR spacer (\textit{i.e.,} larger failure probability $\eta$; green line in Fig.~\ref{fig:resulttwobacteria}b) leads to a higher minimum value for the viral population and weaker oscillations. This could potentially be used to disentangle the effects of growth rate and CRISPR interference on the dynamics.

After a transient period, the dynamics will settle into a stationary state.  The transient is shorter if the  spacer enhanced growth rate $f_1$ is high, or if the failure probability of the spacer $\eta$ is low (Fig.~\ref{fig:resulttwobacteria}, panel a and b). Depending on the choice of initial values and the parameters, there are different steady states.  If spacers are never lost ($\kappa= 0$), we found numerically that  a stable solution occurs when  viruses go extinct and infections cease ($v=0$, $I_{0,1} = 0$).   In this case, the total number of bacteria becomes stationary by reaching capacity ($n=K$), which can only happen when the spacer is sufficiently effective ($\eta<1/b$).  Otherwise bacteria go extinct first ($n=0$) and then the virus persists stably.

  A more interesting scenario occurs when spacers can be lost ($\kappa\neq 0$).  In this case coexistence of bacteria and virus ($n>0$ and $v>0$) becomes possible (see SI for an analytic derivation).   In this case, the bacteria cannot reach full capacity at  steady state---we write $n=K(1-\mathcal{F})$, where the factor
\begin{equation}
	\label{eq:Fdef}
	\mathcal F = 1 - \frac nK
\end{equation}
represents the fraction of unused capacity.  The general expression for $\mathcal{F}$ is given in the SI, and simplifies when the wild type and spacer enhanced bacteria have the same growth rate ($f_1=f_0$) to
\begin{equation}
	\label{eq:single_proto_f_factor}
	\mathcal F = \frac {\kappa} {f_0} \frac {b (1-\alpha) - 1} {(b - 1)(1 - b\eta)} \,.
\end{equation}
Fig.~\ref{fig:resulttwobacteria}c  shows the dependence of $\mathcal{F}$ on the failure probability of the spacer ($\eta$) multiplied by the burst factor ($b$).   We see that even if the spacer is perfect ($\eta = 0$) the steady state bacterial population is less than capacity ($\mathcal{F} > 0$).
These equations are valid when $\mathcal{F} < 1$---this is only possible if the spacer failure probability ($\eta$) is smaller than a critical value ($\eta_c$), 
where
\begin{equation}
	\label{eq:constr2}
	\eta_c= \frac{1}{b} \left(1-\frac{ \kappa}{f_0} \frac{b(1-\alpha)-1}{b-1}\right) + O\left(\frac{r-1}{b^2}\right) \, ,
\end{equation}
where as before $r = f_1/f_0$. This coexistence phase has been found in experiments \cite{Levin2013} where the bacterial population reaches a maximum that is ``phage" limited like in our model. 

In the coexistence phase, the wild type persists at steady state, as observed in experiments \cite{Levin2013}.  In our model, the ratio of spacer-enhanced and wild-type bacteria is 
\begin{equation}
\label{eq:enhanced_to_wild_ratio}
\frac{n_1}{n_0}= \frac{b(1-\alpha)-1}{1- b \eta} \, .
\end{equation}  
This ratio does not depend on the growth rates of the two types of bacteria ($f_1$ \textit{vs.} $f_0$).  So, given knowledge of the burst size $b$ upon lysis,  the population ratio in Eq.~\eqref{eq:enhanced_to_wild_ratio} gives a constraint relating the spacer acquisition probability $\alpha$ and the spacer failure probability $\eta$.  Thus, in an experiment where phage are introduced in a well mixed population of wild type and spacer enhanced bacteria, Eq.~\eqref{eq:enhanced_to_wild_ratio} presents a way of measuring the effectiveness of a spacer, provided the machinery for acquisition of additional spacers is disabled ($\alpha=0$) (\textit{e.g.,} by removing specific \textit{Cas} proteins)~\cite{Barrangou2007, Horvath2010}. Plugging the effectiveness values measured in this way into our model could then be used to predict the outcome of viral infections in bacterial colonies where individuals have different spacers, or have the possibility of acquiring CRISPR immunity.

The lysis timescale $1/\mu$ for infected cells affects the duration of the transient behavior of the population, as  described above.    The longer this timescale, the longer it takes to reach the steady state.  However, the actual size of the steady state population is not dependent on $\mu$ because this parameter controls how long an infected cell persists, but not how likely it is to survive.  This is analyzed in more detail in S1 file.


In previous models, coexistence of bacteria and phage was achieved by hypothesizing the existence of a product of phage replication that specifically affects spacer-enhanced bacteria compared to wild type~\cite{Levin2013}. Here we showed that coexistence is obtained more simply if bacteria can lose spacers, a phenomenon that has been observed experimentally~\cite{Jiang2013, Tyson2008}. More specifically, in our model coexistence requires two conditions: (1)  spacer loss ($\kappa>0$), and (2)  the failure probability of spacers is smaller than a critical value ($\eta<\eta_c$). Our model also reproduces an effect observed by~\cite{Levin2013}, namely that the steady state bacterial population is reduced by the presence of virus.  While this may seem intuitive, previous population dynamics models have not reproduced this finding, which depends critically in our model on the rate of spacer loss.

\subsection*{Effectiveness versus acquisition from multiple spacers}

We can now proceed to analyze the case where multiple protospacers are presented. As before, when we analyze the multiple spacer model, the most interesting case is when the virus and bacteria can co-exist.  The bacteria do not generally fill their capacity when this happens.  The fraction of unused capacity ($\mathcal F = 1 - n/K$) can be characterized using the average failure probability ($\bar\eta$):
\begin{equation}
	\label{eq:multi_calF}
	\begin{split}
		\mathcal F & =\frac {\kappa} {f} \frac {b(1-\alpha) - 1} {(1-b\bar\eta )(b-1)}\, , \\
		\bar \eta &= \frac {\sum_{i=1}^N \eta_i n_i} {\sum_{i=1}^N n_i} \, .
	\end{split}
\end{equation}
\begin{figure}
\begin{center}
\includegraphics[width=1\columnwidth]{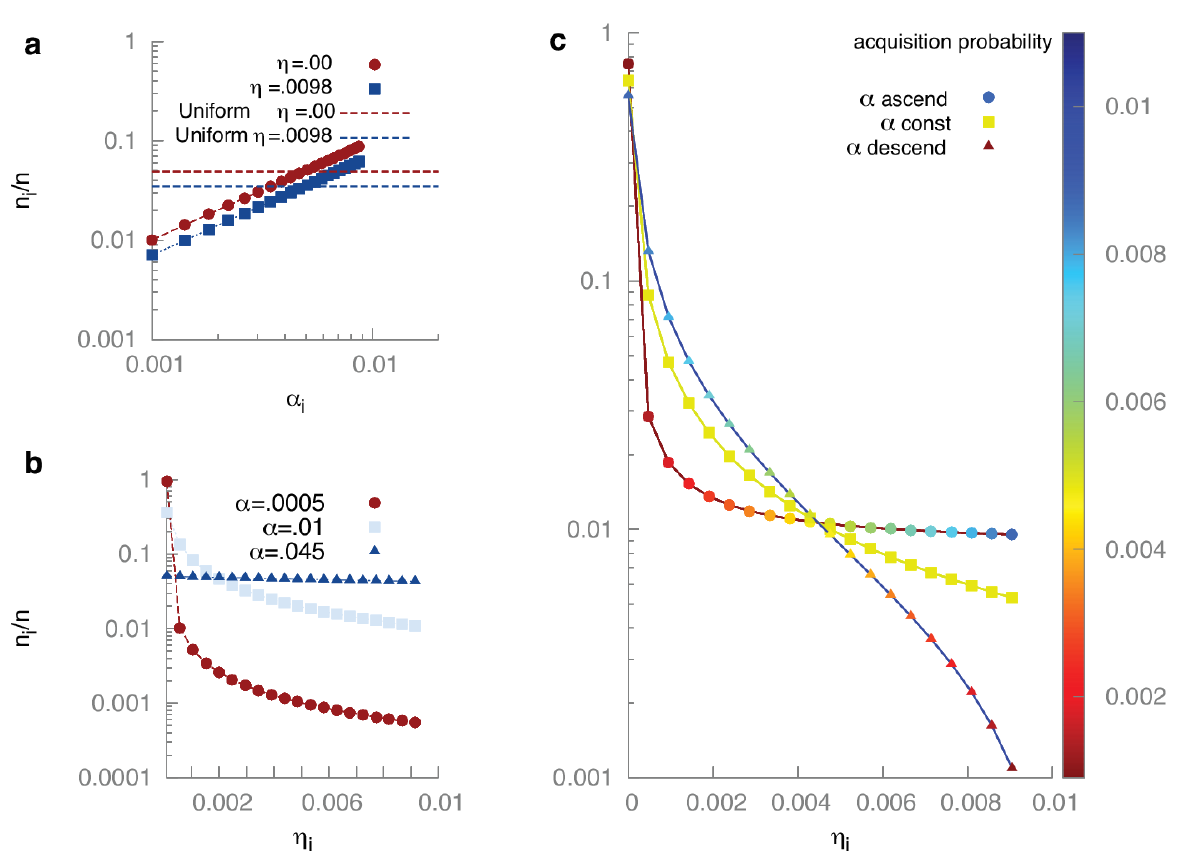}
\end{center}
\caption{The distribution of bacteria with $20$  spacer types.   In these simulations, 100 phage are released upon lysis (burst size $b=100$) and the carrying capacity for bacteria is $K=10^5$.   All rates are measured in units of the bacterial growth rate $f$: the lysis rate is $\mu/f = 1$, the phage adsorption rate is $g/f=10^{-4}$, the spacer loss rate is $\kappa/f=10^{-2}$.
(Panel \textbf{a}) Distribution of spacers as a function of acquisition probability $\alpha_i$ given a constant failure probability $\eta_i=\eta$.   Eq. \eqref{eq:multi_steady_state} shows that the abundance depends linearly on the acquisition probability:  $n_i/n\propto \alpha_i/\alpha$.  Horizontal lines give the reference population fraction of all spacers if they all have the same acquisition probability with the indicated failure probability $\eta$.
(Panel \textbf{b}) Distribution of bacteria with different spacers as a function of failure probability $\eta_i$ given a constant acquisition probability $\alpha_i=\alpha$. For small $\alpha$, the distribution is highly peaked around the best spacer while for large $\alpha$ it becomes more uniform. (Panel \textbf{c})  The distribution of  spacers when both the acquisition probability $\alpha_i$ and the failure probability $\eta_i$ vary. The three curves have the same overall acquisition rate $\alpha=\sum_i\alpha_i=.0972$.   The color of the dots indicates the acquisition probability and the $x$-axis indicates the failure probability of each spacer.  When the acquisition probability is constant (green curve i.e. $\alpha_i=\alpha/20$) the population fraction of a spacer is determined by its failure probability.    If the acquisition probability is anti-correlated with the failure probability (blue curve), effective spacers are also more likely to be acquired and this skews the distribution of spacers even further.  If the acquisition probability is positively correlated with the failure probability (red curve), more effective spacers are less likely to be acquired.  Despite this we see that the most effective spacer still dominates in the population.\label{fig:multispacer}}
\end{figure}%
Bacteria and phage co-exist if $\mathcal F  <  1$ so that $ b \bar \eta < 1-\frac{\kappa (b (1 - \alpha)-1)}{f(b-1)}$.   This is an implicit expression because $\bar\eta$ itself depends on the distribution of bacteria with different spacers.  The coexistence solution can be computed analytically
\begin{equation}
	\label{eq:multi_steady_state}
	\begin{split}
		&g v = b f \mathcal F\,,\\
		&\frac {n_i} {n_0} = \alpha_i \,\frac {b f \mathcal F} {\kappa - f \mathcal F (1 - \eta_i b)}\,,\\
		&\frac{\sum_{i=1}^N n_i }{n_0}= \frac {b(1-\alpha) -1} {1-b\bar\eta}\,.\\
	\end{split}
\end{equation}
We see that the spacer distribution depends on the acquisition and failure probabilities ($\alpha_i$ and $\eta_i$).  As discussed in the single spacer case,
the third equation gives a way to measure the average failure probability ($\bar{\eta}$) of spacers by turning off the acquisition machinery after a diverse population of spacers is acquired~\cite{Barrangou2007, Horvath2010}. (This remains true even if the spacer also affects the growth rate---see S1 file).  Given knowledge of the spacer failure probabilities ($\eta_i$) from single spacer experiments, we can also obtain the acquisition probabilities ($\alpha_i$) by measuring the ratio of spacer enhanced to wild type bacteria ($n_i/n_0$) and using the second equation in \eqref{eq:multi_steady_state}.

The second equation in~\eqref{eq:multi_steady_state} also allows us to make  qualitative predictions about  mechanisms affecting the steady state spacer distribution.  First,  the steady state abundance of each spacer type is proportional to its probability of acquisition ($\alpha_i$). This implies that, if all else is kept fixed, a large difference in abundance can only come from a large difference in acquisition probability (see Fig.\ \ref{fig:multispacer}a).

In contrast, the dependence on the failure probability ($\eta_i$) appears in the denominator,
so that large variations in abundance can follow from even modest differences in effectiveness (Fig.~\ref{fig:multispacer}b). When spacers differ in both acquisition and failure probability, the shape of the distribution is controlled mostly by the differences in effectiveness, with acquisition probability playing a secondary role (Fig.\ \ref{fig:multispacer}c). This suggests that the distribution of spacers observed in experiments, with a few spacer types being much more abundant than the others~\cite{Paez-Espino2013}, is likely indicative of differences in the effectiveness of these spacers, rather than in their ease of acquisition. The distribution of spacers as a function of ease of acquisition and effectiveness is shown for a larger number of spacers in S1 File (Fig.~D), with the same qualitative findings.

Our model also predicts that the overall acquisition probability ($\alpha$) is important for controlling the shape of the spacer distribution. Large acquisition probabilities tend to flatten the distribution, leading to highly diverse bacterial populations, while smaller acquisition probabilities allow the most effective spacers to take over (Fig.\ \ref{fig:multispacer}b). This raises the possibility that the overall spacer acquisition probability of bacteria could be under evolutionary selection pressure as a means of trading off the benefits conferred by diversity in dealing with an  open environment against the benefits of specificity in combatting immediate threats.  This idea could be tested in directed evolution experiments where bacteria are grown in artificial environments with less or more variability in the phage population.

\section*{Discussion}
The CRISPR mechanism in bacteria is an exciting emerging arena for the study of the dynamics of adaptive immunity. 
Recent theoretical work has explored the co-evolution of bacteria and phage~\cite{Levin2013, Han2013, Childs2012}. For example, Levin et al.~\cite{Levin2013} modeled several iterations of an evolutionary arms race in which bacteria become immune to phage by acquiring spacers, and the viral population escapes by mutation. Han et al.~\cite{Han2013} studied coevolution in a population dynamics model in which there are several viral strains, each presenting a single protospacer modeled by a short bit string. Childs et al.~\cite{Childs2012} also used a population dynamics model to study the long-term co-evolution of bacteria and phage. In their model, bacteria can have multiple spacers and viruses can have multiple protospacers, and undergo mutations. 

Our goal has been to model the effect of different properties of the spacers, such as their ease of acquisition and effectiveness, on their abundance in a setting where there is only enough time to acquire a single spacer.  The reason for the latter restriction is that it leads to a more easily interpretable experimental setting. Our goal is not to study long-term bacteria-virus co-evolution, but rather to build a model of the early dynamics of CRISPR immunity that will allow experimentalists to extract key dynamical parameters from their data. An advantage of our model is that it allows study of regimes with a large number of spacer types. We aimed for a  model with the minimal interactions that could explain existing observations, such as an over-abundance of a small number of spacers compared to the rest and the coexistence of phage and bacteria~\cite{Paez-Espino2013, Levin2013, Andersson2008}.   We are specifically interested in the possibility that encounters with a single phage could lead to the acquisition of diverse spacers~\cite{Childs2014}, a phenomenon that could not be explained by the model of Han et al.~\cite{Han2013}.  Likewise, Levin et al.~\cite{Levin2013} did not explicitly model the spacer types and hence could not address their diversity.   Furthermore, their model captured coexistence  by postulating  an as-yet-undetected lysis product from wild type bacteria that harms spacer enhanced ones.   By contrast, we showed above that coexistence, in absence of any other mechanisms of immunity, can be obtained  simply by including spacer loss, which has been experimentally observed \cite{Jiang2013, Deveau2008, Garrett2011}.

Coexistence was also addressed by Haerter el al.~\cite{Haerter2011} and  Iranzo et al.~\cite{Iranzo2013}. Haerter et al.\ exploit  spatial heterogeneity, while our model shows that coexistence can also occur in well-mixed populations. Coexistence in~\cite{Iranzo2013} occurs due to innate immunity for wild type bacteria.  In the latter model, the CRISPR mechanism is taken to incur a cost to the bacteria, and thus loss of the CRISPR locus can occur as a consequence of competition between CRISPR and other forms of immunity, but is not an essential ingredient for coexistence. Their study also focused on longer timescales compared to our work. 
Childs et al.~\cite{Childs2012} discuss the possibility of coexistence, but only in the context of homogeneous bacterial populations, that are either all immune or all wild type. We show that coexistence of both immune and wild type bacteria with phage is possible given a nonzero rate of spacer loss. Finally, Weinberger et al.~\cite{Weinberger2012} used a population genetic model in which the sizes of the bacterial and phage populations are fixed, thus  precluding study of the conditions required for coexistence. The model also did not consider potential differences in the efficacy of spacers. 

Coexistence can also be obtained by placing the bacteria and phage in a chemostat or subjecting them to serial dilutions \cite{Levin1977}. While in some ways this may be a better approximation for natural environments, in this work we focus on experimental conditions in which the interaction takes place in a closed environment. 
We predict that when dilution is negligible, spacer loss is necessary for the existence of a phase where wild-type bacteria, spacer-enhanced bacteria, and phage co-exist.   When there is dilution, coexistence can occur without spacer loss \cite{Levin1977}, but we show in S1 file that this requires a difference in the growth rates of wild-type and spacer-enhanced bacteria.   This difference is known to be small in general \cite{Paez-Espino2013,Jiang2013}, and hence the dilution mechanism for coexistence will lead to  small viral populations at steady state which will be at risk of extinction due to stochastic variation. By contrast, coexistence through spacer loss can support robust steady state viral populations.

We have also addressed factors that influence the spacer distribution across the bacterial population.
This issue was also studied  in He et al.~\cite{He2010} and Han et al.~\cite{Han2013}, but they focused on the way in which diversity depends on position within the CRISPR locus as opposed to the properties of spacers that influence their relative abundance.    Childs et al.~\cite{Childs2014, Childs2012} were also interested in spacer diversity, but assumed that all spacers have similar acquisition probabilities and effectiveness, while we have sought precisely to understand how differences in these parameters affect  diversity.

Our model makes several predictions that can be subjected to experimental test.  First, spacer loss \cite{Jiang2013, Deveau2008, Garrett2011} is a very simple mechanism that allows for coexistence of bacteria and phage. In particular, spacer loss allows coexistence even in the absence of dilution, and permits robust steady state viral populations even if the growth rates of wild-type and spacer-enhanced bacteria are similar.   Direct measurements of the rates of spacer loss may be possible, and would furnish an immediate test of our model.  Alternatively, our model provides a framework for an indirect measurement of the spacer loss rate.   Specifically, this rate is proportional to the viral population and the fraction of unused capacity at steady state.
When the probability of spacer loss is small, our formalism predicts a correspondingly small average viral population.  

Of course, the population in any given experiment experiences fluctuations which could lead to extinction.   An interesting avenue for future work is to include such stochasticity, which would then predict the typical time-scale for viral extinction corresponding to a given probability of spacer loss. This time-scale can be compared with experimental observations~\cite{Paez-Espino2015}. A stochastic model of this dynamics was used by Iranzo et al.~\cite{Iranzo2013}, but did not consider differences in spacer effectiveness.  In order to check whether  the result from a stochastic scenario would be  different from what we found, we checked the stability of the deterministic solution with respect to initial conditions. The system is able to equilibrate in a reasonable time-scale suggesting that the deterministic solution is stable. This is an indication of robustness against stochastic fluctuations.

The effectiveness parameters in our model could be extracted in experiments where bacteria are engineered to have specific spacers~\cite{Bikard2012} and acquisition is disabled \cite{Barrangou2007, Horvath2010}.  In principle the acquisition parameters could be measured by freezing bacterial populations soon after an infection, although initial conditions would require careful control.   Once these parameters are measured, they can be plugged back into the full set of equations to make predictions for the CRISPR dynamics even in the case when acquisition is enabled. A comparison between the measured and predicted dynamics in the presence of CRISPR acquisition would constitute a test of our model.   Alternatively, our model could be fit to measured dynamics to extract the parameters and then tested by comparing with the steady state.

When multiple protospacers are available, we showed that the acquisition probability linearly affects the steady state spacer distribution, while the proportion of more effective spacers is magnified by the dynamics.   Thus, a highly peaked spacer distribution as seen in some studies \cite{Paez-Espino2013} is more likely to occur because of differences in effectiveness if protospacers are acquired with roughly equal probabilities.  In fact, it does seem that some genomic sequences are acquired more frequently than others~\cite{Heler2014,Levy2015}.   While the mechanism for this enhancement has not been fully clarified, one possibility is that the more commonly acquired sequences are simply those that are less prone to mutation in the viral genome. This could be tested by sequencing the virus together with the CRISPR-cassettes in a co-evolving population of bacteria and phage.   This mechanism for enhancing acquisition probability of some spacers is readily incorporated in our model.

Various extensions of our model are possible.  For example, in describing longer timescale experiments we can include the fact that CRISPR cassettes can contain many spacers \cite{He2010}.  Furthermore, we could include the possibility of ``priming'' where the presence of some spacers increases the probability of acquiring others~\cite{Heler2014}.    Such an effect would introduce correlations between different spacer populations $n_i$ and $n_j$ that can be tested experimentally.

Our model showed that high acquisition probabilities will lead to greater diversity in the spacer distribution, while strong selection will tend to homogenize the population of spacers in favor of the most effective one for the current threat.   This suggests that bacteria should adapt the overall spacer acquisition probability to the amount of viral diversity in their environment, perhaps by transcriptional regulation of the {\it cas} genes.    Given an appropriate fitness function and viral landscape our modeling framework could be used to predict the optimal acquisition probability.



\section*{Supporting Information}


\paragraph*{S1 File.}
\label{S1_File}
{\bf Supporting Information.} 

\section*{Acknowledgments}

We thank L.~Marraffini, R.~Heler and members of the Marraffini Lab for help and fruitful interactions.  We also thank  M.~Magnasco, E.~Siggia, B.~Shraiman, C.~Modes, and G.~Falkovich for discussions.   








\end{document}


\title{Supplementary Information\vspace{1em}\\\large Dynamics of adaptive immunity against phage in bacterial populations}

\date{}

\maketitle
\tableofcontents

\renewcommand{\tablename}{Table}
\renewcommand{\figurename}{Fig}

\renewcommand\thefigure{\Alph{figure}}

\section{Materials and Methods}

We numerically integrated our population dynamics equations using custom code written in C.    The order of magnitude of the burst size $b$,  carrying capacity $K$ and growth rate $f$ were taken from \cite{Childs2012}, the spacer loss rate $\kappa$ was estimated from \cite{Jiang2013},  and the spacer failure probability $\eta$ and acquisition probability $\alpha$ were variables that we scanned over.   We took  the death rate $\mu$ of infected bacteria to be comparable to the growth rate $f$.  To estimate the order of magnitude of the phage adsorption rate $g$ we used a simple argument based on diffusion. Because of the large difference in size, we approximate phage particles with points and focus on a single bacterial cell, modeled as a perfectly absorbing sphere of radius $R$. Fick's second law, $\dot c = D \Delta c$, can be used to calculate the concentration profile of phage around the bacteria at stationarity, leading to $c(r) = c_p \bigl(1 - \frac Rr\bigr)$, where $c_p$ is the concentration of phage far away from bacteria. Fick's first law then gives us the flux at the sphere, whose integral gives the rate at which phage are absorbed, $4\pi D c_p R$, where $D$ is the diffusion coefficient. Using Einstein's relation $D = \frac {k_B T} {6\pi \eta r_v}$ to estimate the diffusion coefficient (with $\eta$ the dynamic viscosity of the medium and $r_v$ the size of the virus), we get that an estimate for $g$ is
\begin{equation}
	\label{eq:g_estimate}
	g = \frac {2} {3V} \frac {k_b T} {\eta} \frac {R} {r_v}\,,
\end{equation}
where $V$ is the experimental volume, which only appears here because we defined $g$ in terms of particle numbers instead of concentrations. Using $V\sim 1\unit{\mu L}$, $\eta = 8\times 10^{-4} \unit{Pa\cdot s}$ (for water at $30\degree$), $T = 300 \unit{K}$, $R\sim 1\unit{\mu m}$, $r_v\sim 0.1\unit{\mu m}$, we get  $g \sim 10^{-4} \unit{hour^{-1}}$ . This is very close to experimental values observed in in vitro experiments of bacteria and phage~\cite{Weld2004}.

\section{Detailed dynamics and steady state with one type of spacer}

\subsection{Transient behavior}
In addition to the numerical studies presented in the main text, some aspects of the dynamics of our model can be described analytically. This allows us to get further insights into general features of the solutions and how these features depend on the parameters.

Let us analyze the initial trend in the bacterial population in the single protospacer model (eq.~\mainref{1} in the main text):
\begin{equation}
	\begin{aligned}
		\dot{n}_0 &= f_0 \left(1- \frac{n}{K}\right) n_0 + \kappa n_1 - g v n_0 \,,\\
		\dot{n}_1 &= f_1 \left(1-\frac{n}{K}\right) n_1 - \kappa n_1   - \eta g v n_1 + \alpha \mu I_0 \,,\\
		\dot{I}_0 &= g v n_0 - \mu I_0\,,\\
		\dot{I}_1 &= \eta g v n_1 - \mu I_1\,,\\
		\dot{v} &= b \left(1-\alpha\right) \mu I_0 + b \mu I_1 -g v (n_0+n_1)\, .
	\end{aligned}
	\label{eq:si:single_proto_spacer}
\end{equation}
At the time of inoculation, the population is entirely wild type, so $n_1(0) = 0$, and there are no infected bacteria, $I_0(0) = 0$, $I_1(0) = 0$. The total bacterial population is $n(0) \equiv n_0(0) < K$. The dynamics of $n(t)$ are governed by eq.~\mainref{2} from the main text,
\begin{equation}
	\label{eq:si:single_proto_total}
	\dot n= f_0( n_0 + r n_1) \left(1-\frac{n}{K}\right) -\mu (1-\alpha) I_0 - \mu I_1\,,
\end{equation}
which implies that in this case $\dot n(0) > 0$, \textit{i.e.,} the bacteria always start off with growth. This is a result of the latency between viral infection and death. However, we expect this growth to be short lived: after a time of order $\mu^{-1}$, the infected cells should start dying and the bacterial population should go down. It turns out that growth can sometimes last much longer than that, as shown below.

Suppose we have $t \ll 1/\mu$ and, for simplicity, assume $n_0(0) \ll K$; from the system of equations~\eqref{eq:si:single_proto_spacer} we get
\begin{equation}
	\label{eq:single_proto_short_time_wildtype}
	\begin{split}
		n_0 &= n_0(0) + \bigl[f_0 - g v(0)\bigr] n_0(0)\, t + \mathcal O(t^2)\,,\\
		I_0 &= g v(0) n_0(0)\, t + \mathcal O(t^2)\,.
	\end{split}
\end{equation}
Similarly $n_1 = \mathcal O(t^2)$ and $I_1 = \mathcal O(t^2)$, implying that
\begin{equation}
	\label{eq:single_proto_short_time_all}
	\dot n = f_0 n_0(0) \left\{1 - t \left[g v(0)\left(1 + \frac {\mu (1-\alpha)} {f_0}\right) - f_0\right]\right\} + \mathcal O(t^2)\,.
\end{equation}
If $g v(0) > f_0$ then growth ends in a time
\begin{equation}
	\label{eq:single_proto_short_time_growth_end}
	\begin{split}
		t_p &= \frac 1 {g v (0) \left(1 + \frac {\mu (1-\alpha)} {f_0}\right) - f_0} < \frac {f_0} {\mu (1-\alpha) g v(0)}\\
			&< \frac 1 {\mu (1-\alpha)} \approx \frac 1 \mu\,,
	\end{split}
\end{equation}
assuming that the acquisition probability $\alpha$ is small.
Conversely, if
\begin{equation}
	\label{eq:single_proto_large_growth}
	g v(0) < \frac {f_0} {1 + \mu (1-\alpha)/f_0}\,,
\end{equation}
the initial growth of the wild type bacteria continues past the initial latency time $t = 1/\mu$.   This prediction of the model can be tested via optical density measurements.  Of course, past this time, the approximations we made above no longer hold; eventually the wild type population will decline, being overtaken by the virus, and the bacteria will go through a bottleneck; see Fig. 3 of the main text. Recovery from this bottleneck is due to CRISPR spacer acquisition.

\subsection{Steady state solutions and stability analysis}

\paragraph{Coexistence solution: $v \neq 0$ and $n \neq 0$.}
The steady state solutions for the system in eq.~\eqref{eq:si:single_proto_spacer} are obtained by setting all the derivatives to zero. Solving the equations for the number of infected bacteria and plugging these into the equation for the viral dynamics, we get
\begin{equation}
	\label{eq:steady_Is}
	\begin{split}
		I_0 &=\frac{g}{\mu} {v n_0 }\,,\\
		I_1&=\frac{g \eta}{\mu}v n_1\,, \\
		n_1&=\frac{b (1-\alpha) -1}{1-b\eta} n_0\,,
	\end{split}
\end{equation}
where we assumed $v \neq 0$.

Since $I_i$, $n_i$, and $v$ are population numbers, they must be non-negative.   From eq.~\eqref{eq:steady_Is} above, this requires that $b\eta < 1$ and $b\alpha < b-1$ (in principle the opposite conditions could also hold, $b\eta > 1$ and $b\alpha > b-1$, implying $\alpha > (b-1)/b$. However, in realistic conditions, $b$ is large (of order 100 or 1000), and so $(b-1)/b \approx 1$, while $\alpha$ is typically much smaller than 1). If these conditions are not met, the co-existence solution is not feasible, and we get either $n_i = I_i = 0$ or $v = 0$. 

The remaining steady state values can be obtained after some tedious but straightforward algebra:
\begin{equation}
	\label{eq:solution}
	\begin{split}
		n&= K(1-\mathcal{F}_r)\,,\\
		\mathcal{F}_r &= \frac {\kappa} {f_0}  \frac {b (1-\alpha) - 1} {(1 - b\eta) \left(b - 1 + (r - 1) \cfrac {b (1-\alpha) - 1} {1-\alpha - \eta}\right)}\,,\\
		g v &= f_0 \mathcal F_r + \kappa \, \frac {b(1-\alpha) - 1} {1 - b\eta}\,,\\
		\frac {n_0} {n}  &=\frac{\mu(1- b \eta)}{b \mu (1-\alpha-\eta) + \left(f_0 \mathcal F_r + \kappa \, \cfrac {b(1-\alpha) - 1} {1 - b\eta}\right)(1 - \eta - \eta b \alpha)}\,,\\
		\frac{n_1}{n}&=\frac{\mu(b(1-\alpha)-1)}{b \mu (1-\alpha-\eta) + \left(f_0 \mathcal F_r + \kappa \, \cfrac {b(1-\alpha) - 1} {1 - b\eta}\right)(1 - \eta - \eta b \alpha)}\,,
	\end{split}
\end{equation}
where we see that the virus concentration at steady state is proportional to the rate of spacer loss ($\kappa$). Here the notation $\mathcal F_r$ is used to emphasize the fact that we allow for different growth rates for wild type and spacer enhanced bacteria.

Compared to the expression from eq.~\mainref{3} in the main text, the fraction of unused capacity ($\mathcal F$) changes when the wild type and spacer enhanced growth rates are unequal. The magnitude of the change obeys
\begin{equation}
	\label{eq:correction_F}
\frac{\mathcal{F}_r}{\mathcal{F}}= \frac 1 {1 + \gamma \, \cfrac {b(1-\alpha) - 1} {(b-1)(1 - \alpha - \eta)}} = 1-\gamma \,\frac{ b(1-\alpha) -1 }{(b-1) (1-\alpha-\eta)} + \mathcal O(\gamma^2)\,,
\end{equation}
where $\gamma=r-1$.


The positivity of $n_0$ and $n_1$  also implies that
 $\mathcal F_r < 1$. This translates into a more stringent condition on the failure probability of the spacer ($\eta$). For bacteria to be able to resist infection and for the co-existence solution to be feasible, we need $\eta < \eta_c$, where the critical failure probability is 
\begin{equation}
	\label{eq:eta_more_precise}
	\eta_c = \frac 1b \left(1 - \frac \kappa {f_0} \frac {b(1-\alpha) - 1} {b-1}\right) + \gamma \,\frac{\kappa}{f_0} \frac{b(1-\alpha) -1}{(b-1) (b-1+\kappa/f_0)} - \gamma^2 \,\frac{\kappa}{f_0} \frac{ b\bigl(b(1-\alpha) -1\bigr)}{(b-1+\kappa/f_0)^3} + \mathcal O (\gamma^3)\,.
\end{equation} 
This gives the next couple of terms in the expansion  in eq.~\mainref{4} in the main text. We write this as an expansion in $\gamma = r-1$ because experiments suggest that differences between the growth rates of wild type and spacer enhanced bacteria are small. 

Another interesting limit to consider is when the burst size $b$ is very large, which is the case for typical viruses. When $b\to\infty$, the product between the burst size and the critical failure probability simplifies to
\begin{equation}
	\label{eq:eta_large_b}
	b \eta_c = 1 - \frac {\kappa (1-\alpha)} {f_1} +\mathcal O(b^{-1})\,.
\end{equation}
This means that, apart from being inversely proportional to the burst factor, the critical failure rate at large $b$ only depends on the rate of spacer loss ($\kappa$), the acquisition probability ($\alpha$), and the growth rate of the spacer enhanced bacteria ($f_1$). In particular, it does not depend on the wild type growth rate.

\begin{figure}[t]
	\includegraphics[width=0.49\textwidth]{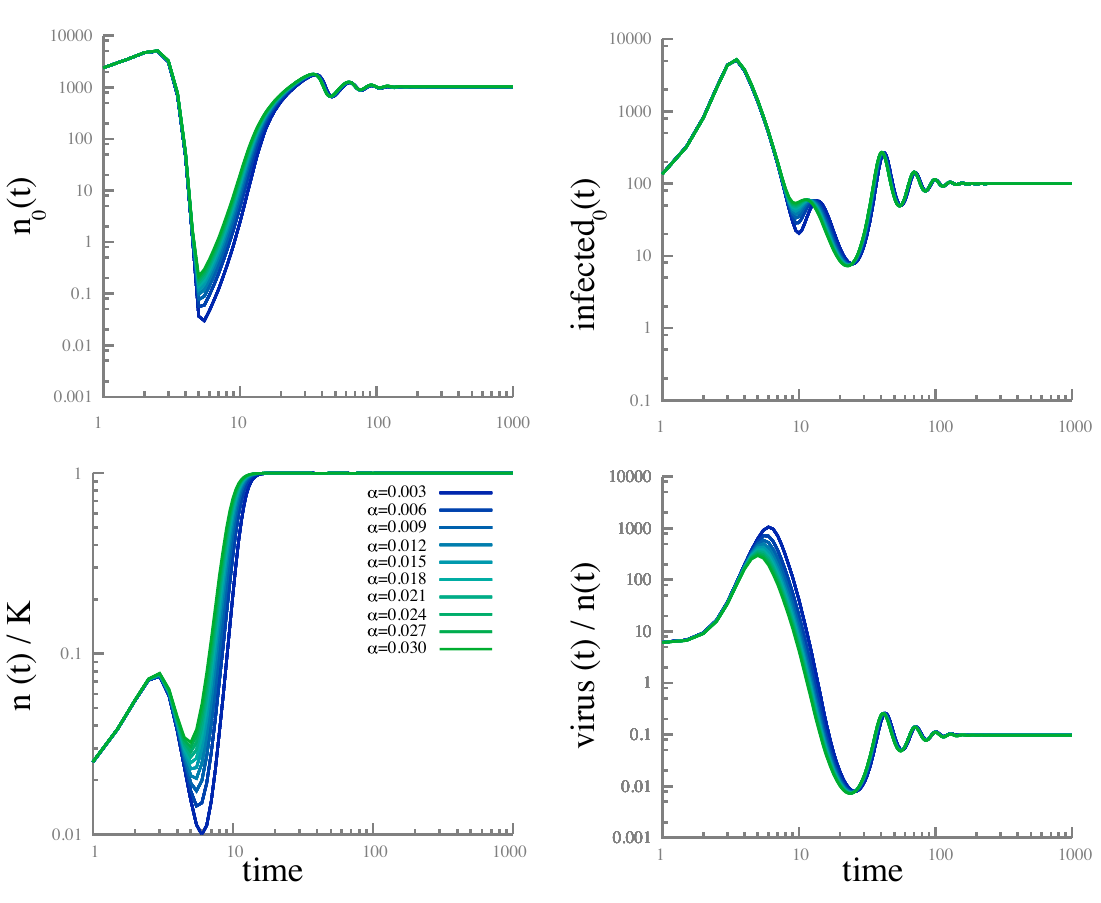}
	\includegraphics[width=0.49\textwidth]{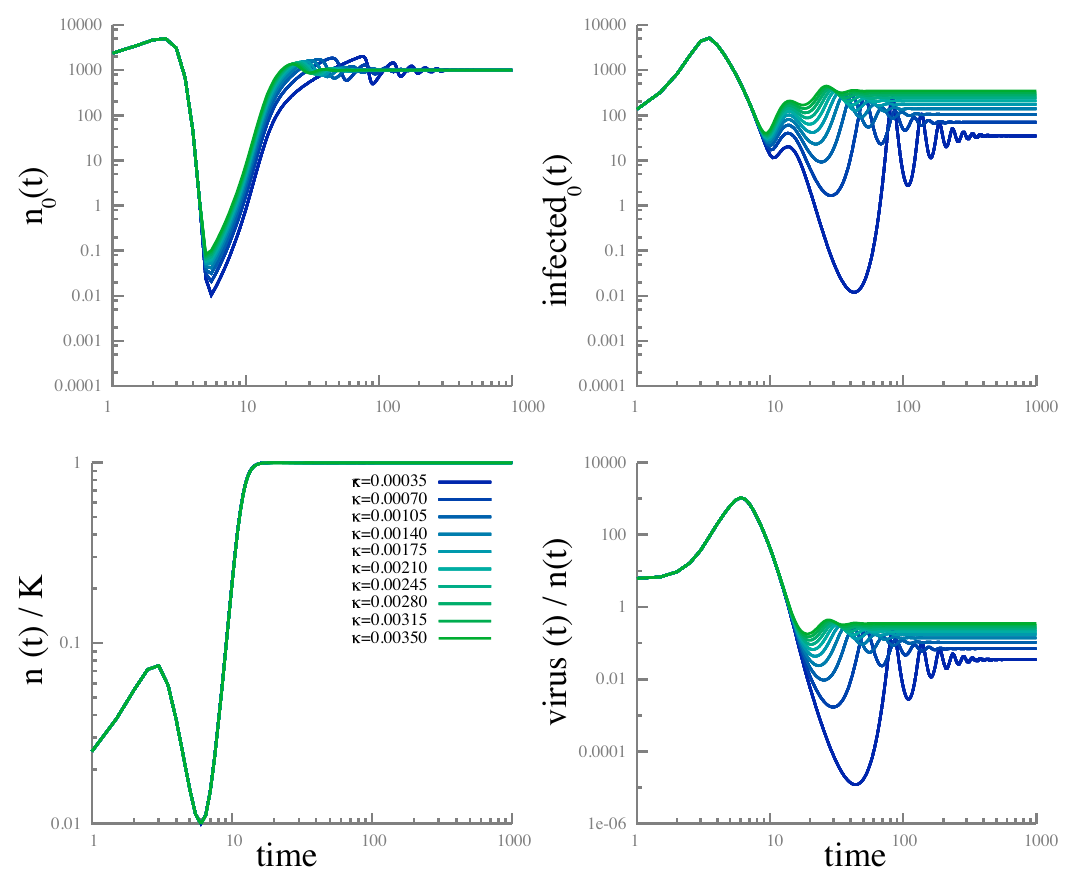}
	\caption{Time-dependence of the population of wild type bacteria, both growing ($n_0$) and infected ($I_0$), the fraction of carrying capacity occupied by bacteria ($n/K$), and the viral population ($v$) for various choices of acquisition probability ($\alpha$; left panel) and rate of spacer loss ($\kappa$; right panel).  Both axes are on a log-scale because the dynamic range varies significantly between the early and late periods.	 Other parameters are chosen similar to the main text: the growth rates of spacer enhanced and wild type bacteria are equal ($r=1$), the phage adsorption rate is given by $g/f_0 = 10^{-5}$, the death rate of infected bacteria is given by $\mu/f_0 = 1$, the burst factor is $b=100$, and the spacer failure probability $\eta=0$. The initial population is all wild type $n=n_0=1000$ and maximum capacity $K=10^5$, and the initial multiplicity of infection (MOI; $v/n$) is 10 such that $v(0)=10^4$. In the left panel, the rate of spacer loss is kept fixed at $\kappa/f_0 = 0.001$. In the right panel, the acquisition probability is kept fixed at $\alpha = 0.003$. We see that the oscillations in the wild type bacteria and viral populations are more strongly damped for large spacer loss rates.\label{fig:parameters}}
\end{figure}
\begin{figure}[tb]
	\center\includegraphics[width=0.49\textwidth]{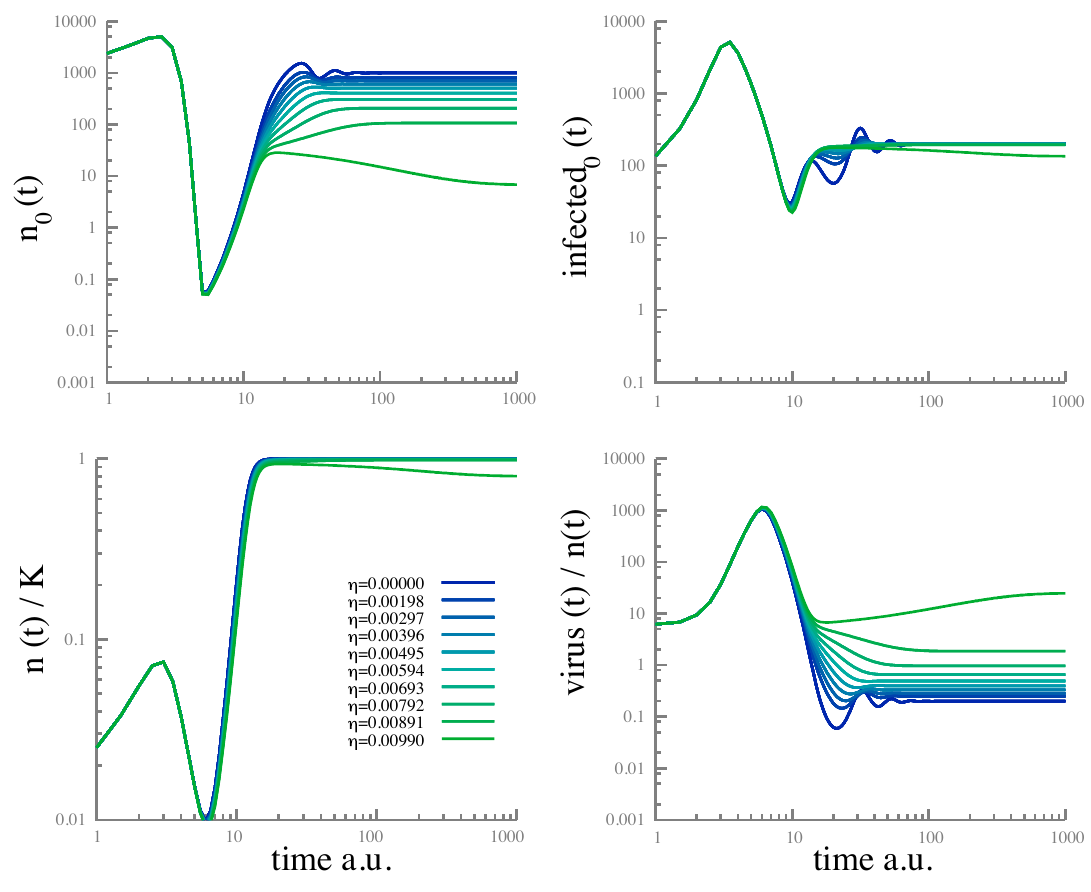}
	\caption{Time-dependence of the population of wild type bacteria, both growing ($n_0$) and infected ($I_0$), the fraction of carrying capacity occupied by bacteria ($n/K$), and the viral population ($v$) for various choices of the spacer failure probability ($\eta$).  Both axes have a log scale to better reveal the dynamical range. Other parameters are chosen similar to the main text: the growth rates of spacer enhanced and wild type bacteria are equal ($r=1$), the phage adsorption rate is given by $g/f_0 = 10^{-5}$, the death rate of infected bacteria is given by $\mu/f_0 = 1$, the burst factor $b=100$, the rate of spacer loss is $\kappa/f_0 = 0.002$ and acquisition probability $\alpha=0.003$. The initial population is all wild type, $n(0)=n_0(0)=10^3$ and the initial multiplicity of infection (MOI; $v/n$) is 10 that means $v(0)=10^4$.\label{fig:eta_dep}}
\end{figure}

A numerical study shows that when the coexistence solution exists ($\eta <\eta_c$), it is also stable for a wide range of parameters. Altering the spacer acquisition probability ($\alpha$), the rate of spacer loss ($\kappa$), or the failure probability ($\eta$) in a wide range does not preclude the coexistence state, although it can lead to significant changes in the population dynamics (see Fig.~\ref{fig:parameters} and Fig.~\ref{fig:eta_dep}). Interestingly, the dynamics of the total number of bacteria ($n$) is almost insensitive to the rate of spacer loss ($\kappa$).
The viral population, and therefore also the population of infected bacteria, are much more strongly affected by changes in the rate of spacer loss. Frequent loss of spacers leads to a stronger damping and shorter period for the oscillations in the viral population, while less frequent loss greatly enhances the amplitude of these oscillations.

The ability to acquire spacers has a large effect on the early dynamics of the bacterial population, while not greatly affecting the viral dynamics. Lower acquisition probabilities lead to a tighter bottleneck for the bacteria, as they require a longer time to gain the CRISPR immunity that they need to fight  viral infection.

The spacer failure probability ($\eta$) does not affect the total bacterial population too much, but has a large effect on the steady state population of phage and wild type bacteria, as well as on the transient dynamics leading to the steady state. More effective spacers lead to larger numbers of wild type bacteria and fewer viruses, shorter transients, and more oscillatory dynamics. In contrast, bacteria with less effective spacers (larger $\eta$) can take a long time to reach steady state, don't seem to exhibit oscillations, and lead to fewer wild type bacteria and more viruses.

When stochastic effects are taken into consideration, random fluctuations could lead to extinction of either bacteria or phage when these go through bottlenecks. Thus, smaller acquisition probabilities lead to higher chances of bacterial extinction, while smaller rates of spacer loss would be dangerous for the phage's survival. Very effective spacers also increase the oscillations in the viral dynamics, while also lowering the steady state viral population, so they too can contribute to viral extinction.


\paragraph{Virus extinction $v=0$.}
When the virus goes extinct ($v=0$), the steady state populations of wild type and spacer enhanced bacteria can depend on the initial conditions.  We can gain some intuition into this case by considering a model in which the virus has already gone extinct; thus, $v=0$ and $I_0 = I_1 = 0$. In this case, the system of equations simplifies to
\begin{equation}
	\label{eq:simplified}
	\begin{split}
		\dot n_0 &= f_0 \left(1- \frac{n}{K}\right) n_0 + \kappa n_1 \,,\\
		\dot n_1 &= f_1 \left(1-\frac{n}{K}\right) n_1 - \kappa n_1 \,,
	\end{split}
\end{equation}
If we assume that the wild type and spacer enhanced growth rates are equal ($f_1=f_0$), the solution can be found analytically:
\begin{equation}
	\label{eq:solution_extinction}
	\begin{split}
&n(t) \ \ = \frac{K n(t_0) \, e^{f_0 (t-t_0)}}{K-n(t_0) \bigl(1-e^{f_0(t-t_0)}\bigr)}\,,\\
&\frac{n_1(t)}{n(t)}= \frac{n_1(t_0)}{n(t_0)} \, e^{-\kappa (t-t_0)}\,,\\
&\frac{n_0(t)}{n(t)}=1-\frac{n_1(t)}{n(t)}\,,
	\end{split}
\end{equation}
where $t_0$ is the initial time and $n(t_0)$ and $n_1(t_0)$ are the initial total population of bacteria and the initial population of spacer enhanced bacteria, respectively. In the long term limit the bacterial population reaches the carrying capacity, $n(t) \to K$. If there is spacer loss ($\kappa\ne 0$), the spacer enhanced bacteria eventually disappear, so the steady state in this case is $n_0 = n = K$, $n_1 = 0$, independently of initial conditions. If there is no spacer loss ($\kappa = 0$), the fraction of bacteria that are spacer enhanced stays constant as the bacteria grow to capacity. 

We can use this result to understand what happens in the more general case when the viral population starts off non-zero but eventually dies out ($v\to 0$). In this case we need to compute the viral extinction time $t_0=t_{\text{eff}}$ such that $v(t_{\text{eff}})\sim 0$.   The arguments above suggest that viral extinction can only happen if there is no spacer loss -- this is because even an exponentially small number of viruses will be able to multiply if presented with wild-type bacteria.   Assuming the spacer loss rate is zero, the fraction of the total bacterial population that contains spacers will stay approximately constant from the time of viral extinction.

\begin{figure}[t]
	\center\includegraphics[width=0.45\textwidth]{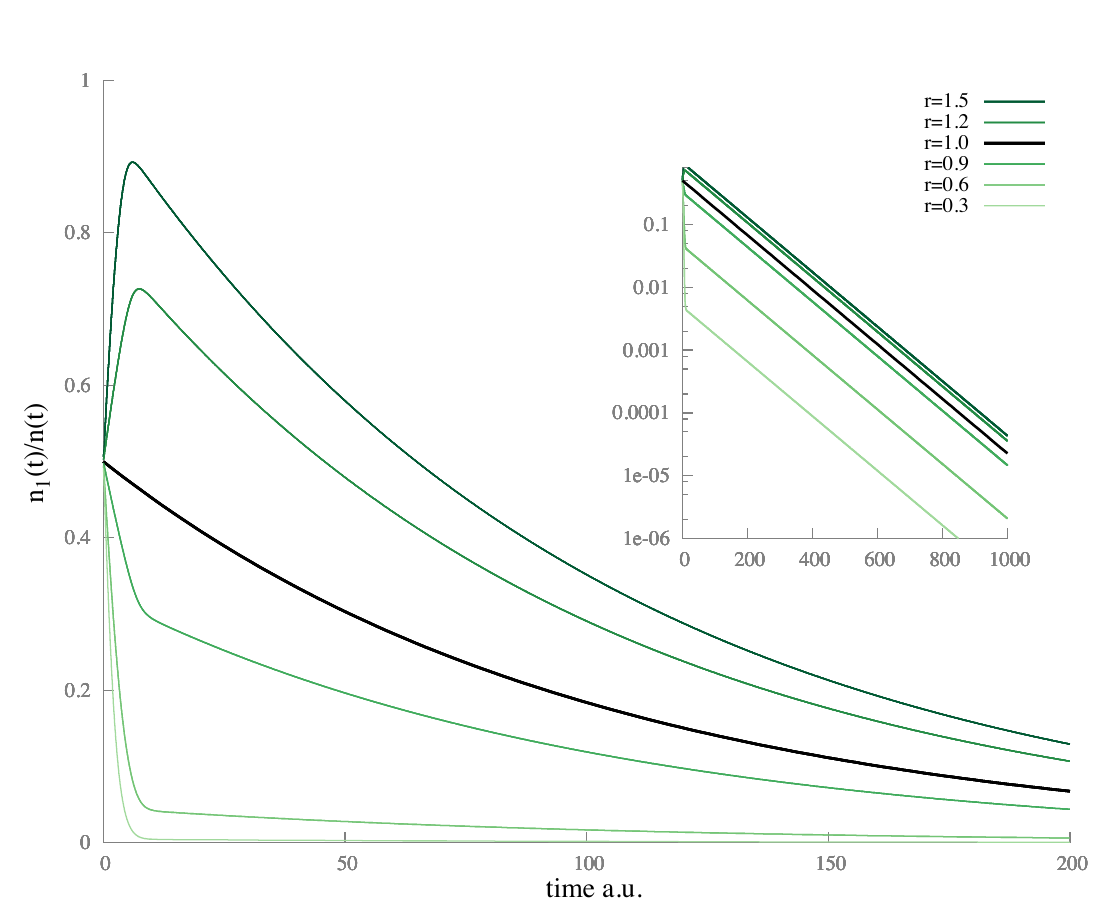}
	\caption{Dynamics of the fraction of spacer enhanced bacteria in absence of virus. The plots show the numerical solution of the system of equations given in (\ref{eq:simplified}) when $K=10^5$ and spacer enhanced and wild type growth rates are different ($r= f_1/f_0 \ne 1$). The initial values $n_1(0)=n_0(0)=50, n(0)=100$ and $\kappa=.01$.  For comparison, we show the analytical solution reported in eq.~\eqref{eq:solution_extinction} for the special case $r=1$ (black line).  The long-term decay of the spacer-enhanced population relative to wild type is simply due to spacer loss in the absence of virus.  In the inset we emphasize the exponential decays for longer time scales by using a log scale.\label{fig:moreratior}}
\end{figure}

We can also investigate what happens when the spacer enhanced growth rate differs from that of the wild type, but this requires numerical simulations (Fig.~\ref{fig:moreratior}). Just like in the case when the growth rates are equal, the population of spacer-enhanced bacteria $n_1(t)$ decays exponentially to zero at large times.

In summary,  when bacteria can lose spacers and the failure probability is lower than the critical value from eq.~\eqref{eq:eta_more_precise}, virus and bacteria co-exist in a steady state after the initial transient dynamics.

\section{Extensions of the model}
In realistic situations, bacteria and viruses undergo decay even in the absence of any external threats. We can incorporate this into our model by adding decay terms into the system of equations~\eqref{eq:si:single_proto_spacer}:
\begin{equation}
	\label{eq:model_with_decay}
	\begin{aligned}
		\dot n_0 &= f_0 \left(1- \frac{n}{K}\right) n_0 + \kappa n_1 - g v n_0  - s n_0\,,\\
		\dot n_1 &=f_1 \left(1-\frac{n}{K}\right) n_1 - \kappa n_1   - \eta g v n_1 + \alpha \mu I_0  -s n_1\,,\\
		\dot I_0 &= g v n_0 - (\mu+s) I_0\,,\\\
		\dot I_1 &= \eta g v n_1 - (\mu+s) I_1\,,\\
		\dot v \ &= b \left(1-\alpha\right) \mu I_0 + b \mu I_1 -g v (n_0+n_1) - s' v\, .
	\end{aligned}
\end{equation}
Here $s$ is the decay rate for bacteria, assumed to be the same regardless whether they are infected or not and whether they are spacer enhanced or wildtype, while $s'$ is the decay rate for phage.

The formalism above can also be used to model a rather different phenomenon: the dynamics of a mixture of bacteria and phage in the case where the experimental preparation includes serial dilutions or chemostats. In these cases, the bacterial culture is either periodically or continuously removed and added to fresh sterile medium. This leads to a continual renewal of nutrients that allows the bacterial population to keep growing.  If natural decay is negligible compared to dilution, we can set $s' = s$ in eqs.~\eqref{eq:model_with_decay}.

To simplify the analysis, we look at two cases: viral decay without bacterial decay, and dilution.  The general case can be treated similarly.

\subsection{Viral decay}
We start with the case in which viruses decay at a rate $s'$, but there is no bacterial decay ($s=0$). We look for steady state solutions for eqs.~\eqref{eq:model_with_decay}, with the added simplification of ignoring spacer loss, $\kappa = 0$.  We will generally assume that the growth rates of spacer enhanced and wild type bacteria are similar, in accordance with experiments. We will thus write $f_1=f_0(1+\gamma)$ and perform an expansion in $\gamma$. These approximations greatly simplify the analytical manipulations without changing the qualitative picture for moderate variations in parameters, as we checked numerically.

There are three types of steady state solutions: 1) a coexistence scenario where spacer enhanced bacteria, wild type  bacteria, and virus coexist; 2) a monoclonal-bacteria scenario where wild type bacteria go extinct leaving only spacer enhanced; and 3) an infection-free scenario where viruses go extinct.

\begin{enumerate}
\item The coexistence scenario.

The steady state solutions expanded to first order in $\gamma=(f_1-f_0)/f_0$ are
\begin{equation}
	\begin{aligned}
	 	n_0 &=  \frac {s' (1-\eta) }{g(b-1) (1-\alpha-\eta) } + \frac {s' \alpha (1 - b \eta)}{g (b-1)^2 (1 - \alpha - \eta)^2} \,\gamma + O(\gamma^2)\,,\\
 		n_1 &=\frac{-s'\alpha}{g(b-1) (1-\alpha -\eta)} + \frac{s' \alpha \bigl(b(1-\alpha)-1\bigr)}{g (b-1)^2 (1 - \alpha - \eta)^2}  \gamma  +O(\gamma^2)\,, \\
 		v \ &=\frac{f_0 \mu  (1-\alpha - \eta) \bigl((b-1) g K-s'\bigr)}{g \bigl[(b-1) g K \mu  (1 - \alpha -\eta) + f_0 s' (1 - \alpha  \eta - \eta)\bigr]} + O(\gamma)\,.
  	\end{aligned}
\end{equation}
Typically, acquisition and failure probabilities are not large, so that we can expect  $\alpha + \eta<1$.  This implies that when the difference in growth rate, $\gamma$, is small, $n_1$ in the solution above is formally negative.  Physically, this means that the coexistence scenario is not feasible in the biologically-plausible region of parameters.

\item The monoclonal-bacteria scenario.

This scenario is characterized by the absence of wild type
\begin{equation}
	n_0=0 \,,\qquad n_1=\frac{s'}{g(b\eta-1)} \,,\qquad v=\frac{f_1 \mu  \bigl[g K(b \eta-1 )-s'\bigr]}{g \eta  \bigl[g K \mu  (b \eta -1)+ f_1 s'\bigr]}\,.
\end{equation}
This solution exists only when the failure rate is high $\eta>1/b$, because $n_1$ must be positive.  This regime is opposite to the one we considered in the main text, where we showed that the bacterial population can survive only when the failure probability $\eta$ is small enough to compensate for the infection bursting factor, $\eta<1/b$. The non-zero viral decay rate acts in favor of the bacteria, giving them a chance to survive even when the failure probability is high.  If spacers can be lost (i.e. $\kappa\ne 0$), then some of the surviving spacer-enhanced bacteria will revert to wild type, thus maintaining a diverse population.  This confirms from a different perspective that spacer loss plays a key role in establishing coexistence of the virus with both the spacer-enhanced and the wild-type bacteria.

\item The infection-free scenario.

In this case, the viral infection is completely cleared, and we get
\begin{equation}
	n_0= K -n_1\,, \qquad n_1=  K \, \frac{b (1-\alpha) -1}{b(1-\alpha-\eta)}-\frac{s'}{b g (1-\alpha-\eta)} \,,\qquad v\to 0\,.
\end{equation}
This is the case where the bacterial population is able to cope with the infection and grows up to maximum capacity. The main difference with respect to the absence of viral decay rate, $s'=0$, is that wild type reaches a higher value proportional to the rate $s'$.
\end{enumerate}

\subsection{The dilution regime}
In the case of either serial dilutions or chemostat conditions, if we ignore natural decay for bacteria and phage, the system dilutes both populations at the same rate, $s' = s$. As above, we start by neglecting spacer loss, $\kappa=0$, and consider conditions in which spacer enhanced and wild type bacteria have almost equal growth rates, so that we can expand in $\gamma = (f_1 - f_0)/f_0$.

We first consider the dynamics in the limit in which the infected bacteria are killed instantaneously, $\mu \to \infty$.   In this case, the system of equations simplifies to
\begin{equation}
	\begin{aligned}
		\dot n_0 &= f_0 \left(1- \frac{n}{K}\right) n_0  - g v n_0  - s n_0\,,\\
		\dot n_1 &= f_1 \left(1-\frac{n}{K}\right) n_1  - \eta g v n_1 + \alpha g v n_0  -s n_1\,,\\	
		\dot v \ &= b \left(1-\alpha\right) g v n_0 + b g v \eta n_1 -g v (n_0+n_1) -s v \,.
	\end{aligned}
\end{equation}
If we consider the case in which infected bacteria do not instantaneously die, the steady state solutions change quantitatively, but not qualitatively. For example, consider the condition on the failure probably $b\eta  = 1$ that defines the boundary of the region where steady state solutions exist.   If the infected bacteria survive for a period of time before dying, it turns out that this condition will be replaced by $b\eta = w$ where $w= \frac{\mu +s } {\mu}$. The dimensionless parameter $w$ effectively rescales the time of phage release from infected cells.

As before, we can identify three classes of steady state solutions.
\begin{enumerate}
\item The coexistence scenario.

Here we get that
\begin{equation}
	\frac{n_1}{n} =  -\frac{\alpha}{1-\alpha -\eta} + \frac{\alpha f_0 ((b-1) g K - s)}{\bigl[f_0 \bigl((b-1) g K-s\bigr)-(b-1) g K s\bigr] (1 - \alpha - \eta)^2}\,\gamma+ O(\gamma^2)\,.
\end{equation}
As above, we expect that $\gamma \ll 1$ and that $\alpha+\eta<1$ -- under these conditions the positivity of $n_1$ implies that the coexistence solution is not feasible.

\item The monoclonal-bacteria scenario.

Here wild type bacteria are absent:
\begin{equation}
	\begin{aligned}
		n_0 &= 0 \,,\\
 	n &= n_1 =\frac{s}{g (b\eta-1)} \,,\\
	g v \eta &= f_1 \left(1-\frac{s}{g K (b \eta -1 )}\right) -s\,.
 	\end{aligned}
\end{equation}
This solution is feasible when $\eta >1/b$, a regime where failure rate is high, opposite to the situation considered in the main text.  The virus only survives because the spacer fails with a relatively high probability.  However, if the dilution is large ($s>g K (b \eta-1)$) there will be no viruses left at steady state because growth of  virus in the bacteria is more than compensated for by loss due to dilution.   Again in this regime adding spacer loss will keep wild type bacteria alive, allowing for coexistence of the two species of bacteria and the virus.

\item Low infection scenario.

The only scenario in which virus and both types of bacteria co-exist without spacer loss is one where  the spacer-enhanced and wild-type bacteria grow at different rates ($\gamma \neq 0$). The steady state solutions at leading order in $\gamma$ are
\begin{equation}
	\begin{aligned}
		\frac{n_1}{n} &= x_1^{(0)} + \gamma x_1^{(1)} + O(\gamma^2)\,,\\
		n_0 &= n - n_1\,,\\
		n &= K\left(1-\frac{s}{f_0}\right) +  \gamma  x_1^{(1)} \frac{b g K^2  (1-\alpha -\eta ) (f_0-s)^2}{f_0^2 s}+O(\gamma^2)\,,\\
		v &= -\frac{ s \bigr[g K(f_0 -s ) \bigl( b (1-\alpha )- 1\bigr) - f_0 s \bigr]}{g (1-\alpha -\eta ) \bigl[g K (b-1)(f_0 - s) - f_0 s\bigr]} \gamma+O(\gamma^2)\,,
  	\end{aligned}
\end{equation}
where the two expansion coefficients for the fraction of spacer enhanced bacteria are
\begin{equation}
	\begin{aligned}
		x_1^{(0)}  &= \frac{  b(1-\alpha ) -1  }{ b (1-\alpha -\eta ) } -\frac{f_0 s}{bg K (f_0-s) (1-\alpha-\eta)}\,,\\
		x_1^{(1)}  &= f_0 s^2 \frac {gK (f_0 - s)\bigl(b(1-\alpha)-1\bigr) - f_0 s} {b g K \bigl[g K (b-1) (f_0-s)- f_0 s\bigr] (1 - \alpha - \eta)^2 (f_0-s)^2 }\,.
	\end{aligned}
\end{equation}
The number of viruses is proportional to the relative difference in growth rates, $\gamma$, so it is different from zero only when spacer enhanced and wild type bacteria grow at different rates. 
  
In order to analyze the parameters where the solution is feasible, we further expand in the dilution parameter $s$, and  get
\begin{equation}
	x_1^{(0)} = \frac{b(1-\alpha)-1}{b(1-\alpha-\eta)} - \frac{s}{b g K (1-\alpha-\eta)} + O(s^2) \,,\qquad x_1^{(1)} =O(s^2)\,.
\end{equation}
This means that for small dilution, the total bacterial population is not resource-limited, but there is a correction due to dilution. The coexistence between bacteria and virus is controlled by the relative difference in growth rate, $\gamma$, and the dilution, $s$:
\begin{equation}
	\begin{aligned}
		n &\approx K\left(1-\frac{s}{f_0} + \gamma s\, \frac {b(1-\alpha) - 1}{ f_0 (b-1) (1-\alpha -\eta)} \right)\,,\\
		n_1 &\approx K \, \frac{b(1-\alpha)-1}{b(1-\alpha-\eta)} - s\,\frac{g K \bigl(b(1-\alpha )-1\bigr)+f_0}{f_0 b  g (1-\alpha -\eta )} + \gamma s \, K\, \frac{\bigl(b(1-\alpha)-1\bigr)^2}{b f_0 (b-1) (1-\alpha -\eta )^2} \,,\\
		v &\approx-\gamma  s\, \frac{ b (1-\alpha)  -1 }{g (b-1) (1-\alpha -\eta )}\,.
 	\end{aligned}
\end{equation}
For this solution to be feasible, we need $v>0$. If the spacer enhanced bacteria have a higher growth rate than the wild type, $\gamma > 0$, this can only happen if $\alpha > 1 - 1/b$, implying an unrealistically high (close to unity) acquisition probability. The more realistic scenario occurs when the spacer enhanced bacteria grow slower than the wild type, $\gamma < 0$. In this case, coexistence between both bacterial species and phage can be observed, but the amount of phage is small, since it is proportional to the product of the relative difference in growth rates and the dilution rate, both of which are typically small.

Notice that the solutions we have found are not fundamentally new.  Rather, they represent small corrections to the solutions we found without dilution.  In the main text we found that coexistence of the phage with both bacterial species was enabled by spacer loss which then also implied that the bacterial population did not reach capacity.  Here we see that dilution provides an alternative mechanism (other than spacer loss)  for these effects, but only if spacer-enhanced and wild-type bacteria grow at different rates ($\gamma \neq 0$).  Since this rate difference is measured to be small, and since dilution in typical experiments is an order of magnitude smaller than growth rate, we can conclude that the latter scenario for coexistence leads to small viral populations.   Stochastic effects are then likely to lead to extinction of the virus.   The spacer-loss mechanism discussed in the main text can lead to more robust viral populations at co-existence.

\end{enumerate}

\section{Detailed steady state solution for multiple spacers}
In the main text, we showed the steady state values for the case of multiple spacers only when all the growth rates were the same. Here we generalize the dynamics from eq.~\mainref{6} in the main text
to the case where each spacer ($i$) has a different growth rate $f_i$,
\begin{equation}
\label{multi_acquisition}
\begin{aligned}
\dot n_0 &=  f_0\left(1- \frac{n}{K}\right) n_0 + \kappa \sum_{i=1}^N n_i - g v n_0\,,\\
\dot n_i &=f_i \left(1-\frac{n}{K}\right) n_i - \kappa n_i  - \eta_i g v n_i + \alpha_i \mu I_0\,, \\
\dot I_0 &= g v n_0 - \mu I_0\,,\\
\dot I_i &= \eta_i g v n_i - \mu I_i\,,\\
\dot v \ &= b \Bigl(1-\sum_{i=1}^N \alpha_i\Bigr)\mu I_0 + b  \mu \sum_{i=1}^N I_i  - g v \Bigl(n_0 + \sum_{i=1}^N  n_i\Bigr)\,.
\end{aligned}
\end{equation}
Setting the time derivatives to zero, we obtain
\begin{equation}
	\label{eq:multisteady_Is}
	\begin{split}
		&I_0 =\frac{g}{\mu} {v n_0 }\,,\\
		&I_i=\frac{g \eta_i}{\mu}v n_i\,, \\
		&\frac{\sum_i n_i}{n_0}=\frac{b (1-\alpha) -1}{1-b\bar\eta}\,,
	\end{split}
\end{equation}
where the average failure probability ($\bar \eta$) is defined as in the main text, $\bar \eta = \sum\eta_i n_i / \sum n_i$ (eq.~\mainref{8}). Similar to the case of a single spacer, we see that the ratio between the total number of spacer enhanced bacteria and the number of wild type bacteria is independent of the growth rates ($f_i$). 
\begin{figure}[htb]
\begin{center}
	\includegraphics[width=0.5\textwidth]{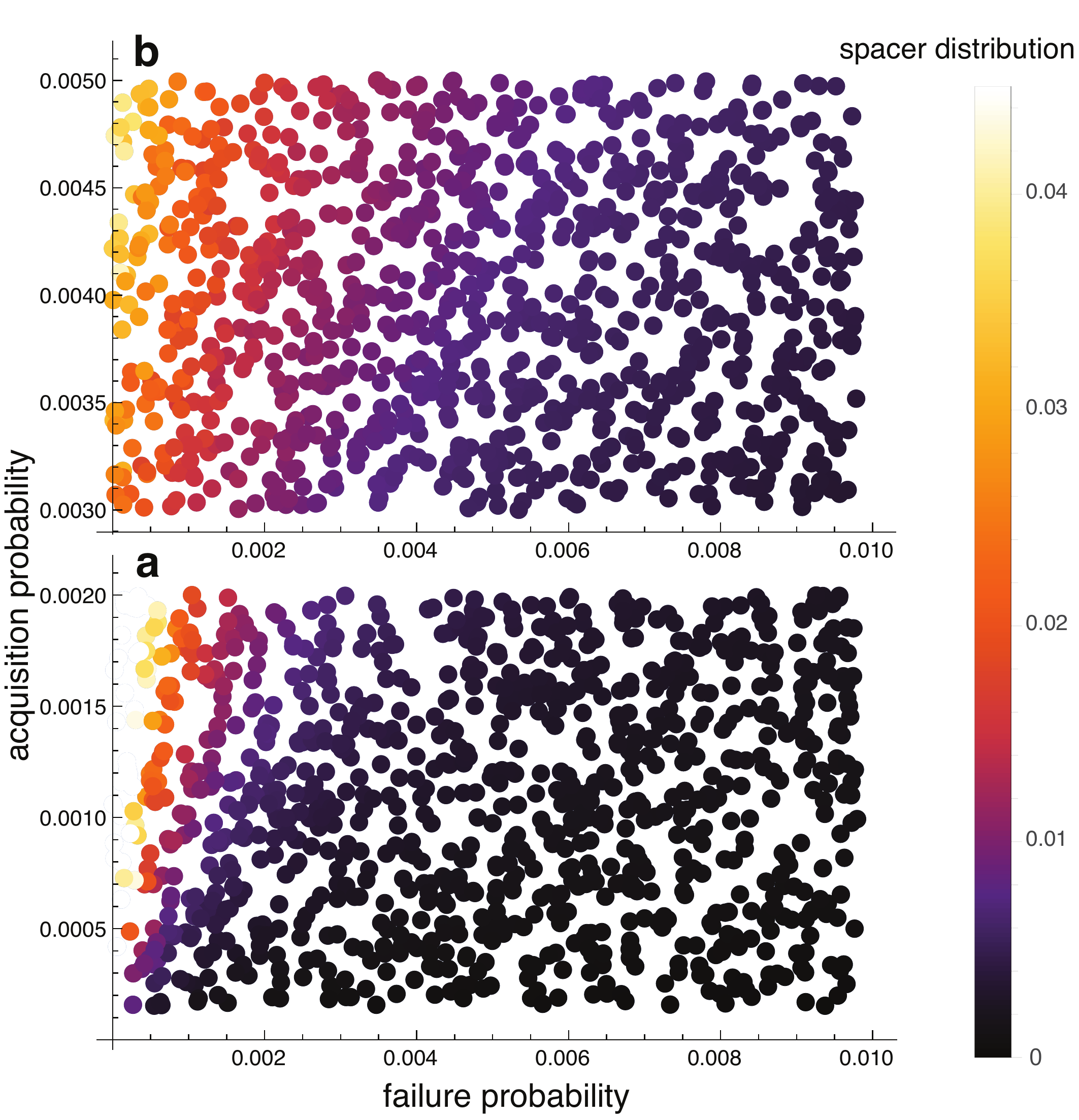}
	\end{center}
	\caption{Distribution of spacers as a function of failure probability ($\eta_i)$ and acquisition probability ($\alpha_i$). The plot was obtained using eqs.~\eqref{eq:full_multispacer} for a set of 100 spacers with failure probability below the critical value $\eta<\eta_c$ and acquisition probabilities chosen uniformly in two different regimes. Panel \textbf{a}) First regime of small overall acquisition probability $\alpha = \sum \alpha_i \approx 0.1$. As explained in the main text, this leads to a very homogenous population where the best spacers are more abundant. There is a strong dependence on the failure probability, which can be seen by the presence of closely spaced almost vertical contour lines. Panel \textbf{b}) Second regime of large overall acquisition probability, $\alpha \approx 0.4$. This tends to reduce the importance of differences in failure probability as shown by contour lines spaced far apart that bend horizontally. \label{fig:alphaEtaMap}}
\end{figure}

By introducing the growth rate ratio ($r_i=f_i/f_0$) and the average growth rate ratio 
\begin{equation}
	\label{eq:avg_defs}
	r = \frac {\sum_{i=1}^N r_i n_i} {\sum_{i=1}^N n_i}\,.
\end{equation}
 we can now obtain the fraction of unused capacity ($\mathcal {F}_r=1-n/K$):
\begin{equation}
	\label{eq:fraction_multispacer}
\mathcal F_r= \frac \kappa {f_0} \frac {b (1-\alpha) - 1} {(1-b \bar\eta) \left(b-1 + (r-1) \cfrac {b(1-\alpha) - 1} {1 - \alpha - \bar\eta}\right)}\,,
\end{equation}
which generalizes eq.~\mainref{8} in the main text. The remaining steady state values are given by:
\begin{equation}
	\label{eq:full_multispacer}
	\begin{split}
		n\ &=K(1-\mathcal F_r)\,,\\
		gv &= f_0 \mathcal F_r + \kappa \frac {b(1-\alpha) - 1} {1 - b \bar \eta}\,,\\
		\frac {n_0} n &= \frac {\mu(1 - b\bar \eta)} {b \mu(1 - \alpha - \bar\eta) + \left(f_0 \mathcal F_r + \kappa \cfrac {b(1-\alpha) - 1} {1 - b\bar \eta}\right)(1 - \bar \eta - b \bar \eta \alpha)}\,,\\
		\frac {n_i} {n_0} &= \alpha_i \, \frac {f_0 \mathcal F_r + \kappa \frac {b(1-\alpha) - 1} {1 - b\bar \eta}} {\kappa \left(1 + \eta_i \cfrac {b(1-\alpha) - 1} {1 - b\bar\eta}\right) - f_0 \mathcal F_r (r_i - \eta_i) }\,.
	\end{split}
\end{equation}
Just as in the case when the growth rates are all equal (eq.~\mainref{9} in the main text), the distribution of spacers shows a linear dependence on the acquisition probability $\alpha_i$.

Fig.~\ref{fig:alphaEtaMap} shows how the fraction of the bacterial population containing a specific spacer ($n_i/n$) depends on that spacer's failure probability ($\eta_i)$ and acquisition probability ($\alpha_i$). This is shown for the case when all the growth rates are equal ($r_i = 1$). Compare this to Fig.~4 in the main text, which gives a different way of looking at these results.
